\documentclass[conference, 10pt]{IEEEtran}

\usepackage[utf8]{inputenc}
\usepackage{amssymb}
\usepackage{pifont}
\usepackage{booktabs}
\usepackage{morefloats}
\usepackage{graphicx}

\usepackage{array}
\usepackage{enumitem}
\usepackage{multirow}
\usepackage{xspace}
\usepackage{fdsymbol}
\usepackage{subfigure}
\usepackage{booktabs}
\usepackage{adjustbox}
\usepackage{wasysym}
\usepackage{hyperref}
\newcommand{\num}{four }

\newcommand{\cmark}{\ding{51}}%
\newcommand{\xmark}{\ding{55}}%
\newcommand{\tild}{$\sim$}%

\newcommand{\Tdot}{$\medblackcircle$}
\newcommand{\TDot}{$\medcircle$}

\newcommand{\Thdot}{$\LEFTcircle$}

\newcolumntype{R}[2]{%
    >{\adjustbox{angle=#1,lap=\width-(#2)}\bgroup}%
    l%
    <{\egroup}%
}

\newcommand*\rota{\multicolumn{1}{R{90}{1em}}}
\newcommand{\etal}{\emph{et al.}\xspace}
\newcommand{\Par}[1]{\smallskip\noindent{\textbf{\textit{ #1.}}}}

\newcommand{\up}{
Synthetic Sensor Spoofing\xspace}

\setlist[itemize]{itemsep=0mm}
\usepackage{paralist}

\begin{document}

\title{No Need to Know Physics:\\ Resilience of Process-based Model-free Anomaly Detection for Industrial Control Systems}
\author{
    \IEEEauthorblockN{Alessandro Erba, Nils Ole Tippenhauer}
    \IEEEauthorblockA{CISPA Helmholtz Center for Information Security
    \\alessandro.erba@cispa.de, tippenhauer@cispa.de}
}

\maketitle

\begin{abstract}
In recent years, a number of process-based anomaly detection schemes for Industrial Control Systems were proposed. In this work, we provide the first systematic analysis of such schemes, and introduce a taxonomy of properties that are verified by those detection systems. We then present a novel general framework to generate adversarial spoofing signals that violate physical properties of the system, and use the framework to analyze four anomaly detectors published at top security conferences. We find that three of those detectors are susceptible to a number of adversarial manipulations (e.g.,~spoofing with precomputed patterns), which we call Synthetic Sensor Spoofing and one is resilient against our attacks. We investigate the root of its resilience and demonstrate that it comes from the properties that we introduced. Our attacks reduce the Recall (True Positive Rate) of the attacked schemes making them not able to correctly detect anomalies. Thus, the vulnerabilities we discovered in the anomaly detectors show that (despite an original good detection performance), those detectors are not able to reliably learn physical properties of the system. Even attacks that prior work was expected to be resilient against (based on verified properties) were found to be successful. We argue that our findings demonstrate the need for both more complete attacks in datasets, and more critical analysis of process-based anomaly detectors. 
We plan to release our implementation as open-source, together with an extension of two public datasets with a set of \up attacks as generated by our framework.

\end{abstract}

\section{Introduction}
Industrial Control Systems (ICS) enable the control of physical processes by the interaction of computers, communication networks, sensors, and actuators. Sensor data are used to feed a control program that acts on the physical world by sending commands to actuators. Examples of such systems are water distribution systems, manufacturing plants, and smart grids.

Control actions are taken based on sensor data, thus ICS are threatened by sensor spoofing attacks, which maliciously manipulate the information reported by the sensors. A number of strategies to select spoofing values have been proposed, among them stealthy attacks~\cite{urbina16limiting}, Replay attacks~\cite{mo2009secure}, and concealment  attacks~\cite{erba2020concealment}. According to the attack point, sensor spoofing can affect the ICS at different levels. If applied between a sensor and controller, spoofing can directly affect the control action driving the system to anomalous states. If applied after controllers (e.g.,~when transmitted to supervising systems), sensor spoofing can prevent the detection of ongoing manipulations of the physical process. Regardless of the attack point and purpose, we assume attackers aim to remain hidden to increase their impact~\cite{garcia17hey}.

To address those threats, a number of intrusion and anomaly detection schemes for ICS and in general for Cyber-Physical Systems (CPS) have been proposed in the literature~\cite{feng2019systematic, chen2018learning, ahmed18noiseprint, goh2017anomaly, taormina2018deep, kravchik2018detecting, aoudi18truth, urbina16limiting}.  In particular, process-based anomaly detection~\cite{feng2019systematic} schemes leverage actuator and sensor data to detect anomalies in the process and operations of the system. We differentiate \emph{model-free} and \emph{model-based} anomaly detectors, depending on whether a detailed physical model is leveraged by the scheme. The former category considers the physical-process as a black-box, and uses Machine Learning~\cite{chen2018learning, ahmed18noiseprint}, Deep Learning~\cite{goh2017anomaly, taormina2018deep, kravchik2018detecting}, System Identification~\cite{aoudi18truth, urbina16limiting}, and Data Mining~\cite{feng2019systematic} techniques during the training of the classifier/predictor. The latter category treats the physical process as a white-box (e.g.,~by using a set of linear or non-linear equations describing the physics), and uses techniques from control theory for detection~\cite{urbina16limiting, choi2018detecting, quinonezsavior}. Here, we focus on attacks on model-free anomaly detection, as many model-based methods were already found vulnerable~\cite{dash2019out, quinonezsavior, shen2020drift}. Of the model-free anomaly detection systems, only Deep-Learning based schemes were so far attacked~\cite{erba2020concealment}, and we thus focus on the other types in this contribution (Machine Learning, System identification, and Data mining).

In this work, we analyze prior work ICS anomaly detection systems from an adversarial perspective, assuming the attacker wants to evade detection of anomalous system states by spoofing selected sensor values (consistent with the attacker modeled in prior work~\cite{garcia17hey}). Then we characterize the properties of the anomaly detection problem w.r.t. the properties of multivariate time series in ICS. We show how to construct \emph{\up} (e.g.,~Replay attacks, attacks with simulated noise) that:
i) are built without explicit knowledge of the physical model of the system,
ii) respect attacker model assumptions from previous works,
iii) are not detected by most of of the analyzed detection schemes (while detectable by humans).

This demonstrates that (surprisingly) an attacker \emph{does not need to know physics} to evade model-free process-based detectors (in contrast to evading human operators~\cite{garcia17hey}). We then investigate the reason why the tested detection mechanisms fail or succeed to spot \up attacks. 

We summarize our main contributions as follows:
\begin{compactitem}
    \item We provide the first systematic analysis of process-based anomaly detectors and show that they are susceptible to a number of simple adversarial manipulations. 
    \item We present a general framework to create \up attacks. This framework can be used to find vulnerabilities of anomaly detectors that are not 
    appropriately represented in related public datasets. 
    \item Using our framework, we practically implement the attacks, and demonstrate their efficacy against \num model-free anomaly detectors form literature. We show that (surprisingly) even very basic Constant spoofing attacks are effective (e.g.,~leading to a Recall of 0.0).
\end{compactitem}
We plan to release our open-source framework implementation, together with an extension of two public datasets containing \up attacks as generated by our framework.

Our work is structured as follows. In Section~\ref{sec:background} we present background of our work. In Section~\ref{sec:properties} we identify the challenges for anomaly detection and Sensor Spoofing, and provide a taxonomy of related work. In Section~\ref{sec:system} we introduce System and Attacker Model, provide a motivating example, and present the design of our framework. In Section~\ref{sec:framework}, we present our framework for \up. In Section~\ref{sec:evaluation} we present experimental setups. In Section~\ref{sec:results} evaluation results. We summarize related work in Section~\ref{sec:related}, and conclude the manuscript in Section~\ref{sec:conclusion}.

\section{Background}
\label{sec:background}
In this section, we introduce systems in State-space representation an abstraction of an Industrial Control System and then provide an overview of related anomaly detection systems that we will analyze in this work.

\subsection{State-space representation}

Physical systems can be represented with the so called State-space representation~\cite{Chen98linear}, Equation~\ref{eq:state-space} represents a discrete time system. This representation combines the input and the state of the system to derive the evolution of the state and the output of the system.
\begin{equation}
\label{eq:state-space}
\begin{cases}
    x_{k+1} = Ax_k+Bu_k \\
    y_k = Cx_k+Du_k     
\end{cases}
\end{equation}

Where $k := kT$ and $T$ is the sampling time. $x_k \in \mathbb{R}^n$ represents the state of the system, that is defined as the set of variables (directly or indirectly measurable) that characterize the physical system at a given time. This set of variables defines an Euclidean-space i.e.,~the State-space, and the state of the system at time $k$, i.e.,~$x_k$ is a vector in the State-space. 
$u_k\in \mathbb{R}^p$ represents the input (or control vector) to the system, it influences the state of the system $x_k$  and its output $y_k$. In a feedback control loop, $u_k$ is the output of the controller. 
$y_k\in \mathbb{R}^q$ represents the output of the system, and it can be measured with sensors and it is influenced by input $u_k$ to the system in the state $x_k$.
$A\in \mathbb{R}^{n\times n}$ is the state matrix, it contains the coefficients of the dependence between $x_k$ and its derivative. 
$B\in \mathbb{R}^{n\times p}$ is the input matrix, it contains the coefficients of the dependence between $u_k$ and the state derivative. 
$C\in \mathbb{R}^{q\times n}$ is the output matrix, it contains the coefficients of the dependence between $x_k$ and $y_k$. 
$D\in \mathbb{R}^{q\times p}$ is the feed-through matrix, it contains the coefficients of the dependence between $u_k$ and $y_k$. In a time invariant system, $A$, $B$, $C$, $D$ are constants. 

\subsection{Industrial Control Systems}
\begin{figure}
    \centering
    \includegraphics[width=0.7\linewidth]{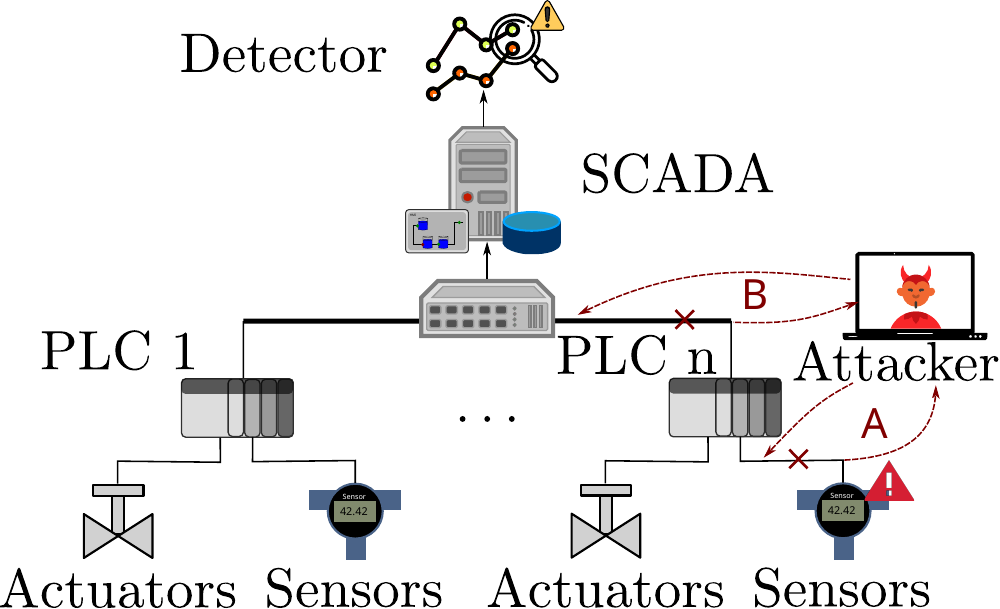}
    \caption{Example Industrial Control System. The PLC reports sensor readings to an SCADA server. The information received by the SCADA is analyzed by the anomaly detector. An attacker can spoof the information transmitted from controller to the historian to conceal its attacks over the physical process}
    \label{fig:system_architecture}
\end{figure}

Industrial Control Systems (ICS) are widely used to automate processes in production plants and facilities, they are an example of systems that can be modeled in the State-space representation. As depicted in Figure~\ref{fig:system_architecture}, ICSs are composed of Cyber and Physical components, interconnected among each other to interact with the physical environment. Cyber components comprise the hardware and software that are used to control the process. In an ICS, Programmable Logic Controllers (PLC) are deployed to implement the system's control logic. PLCs observe sensor values and send commands to actuators. Moreover, sensor data and actuator state are sent to a Supervisory Control and Data Acquisition System (SCADA), for monitoring and historian purposes. ICS components communicate using proprietary communication protocols~\cite{galloway2013introduction}. Due to legacy compliance and resource constraints, industrial protocols typically do not feature any security mechanisms such as authentication or encryption.

\subsection{Detection Schemes and Datasets}
\label{sec:schemes}
Different detection process-based schemes have been proposed to solve the anomaly detection problem in ICS~\cite{feng2019systematic, chen2018learning, ahmed18noiseprint, goh2017anomaly, taormina2018deep, kravchik2018detecting, aoudi18truth, urbina16limiting}. Those solutions leverage the system sensor readings to spot the presence of ongoing anomalies. Sensor readings are analyzed by the process-based anomaly detector to extract an anomaly score, this anomaly score is then compared to a threshold. If the threshold is surpassed, an alarm is raised. The anomaly score can be computed in different ways. We call \emph{Model-free} detectors, the detectors that compute the anomaly score based on a black-box model of the process generating sensor data. This reduces the engineering effort to model system behavior and increases the detector's generality. We call \emph{Model-based} detectors, the detectors that compute the anomaly score according to a white-box knowledge of the physical process that generates the sensor data, e.g.,~the State-space representation of the process. This requires ad-hoc modeling of the system behavior (see Section~\ref{sec:taxonomy} for the detailed description of related work).

Datasets to evaluate detection schemes are collections of sensor readings and actuator states recorded from ICS testbed or process simulation~\cite{taormina18battle, wadi17dataset, mathur16swat, goh2016dataset}. Data are usually organized into two data captures. The first `train dataset' is recorded during normal operating conditions, the second `test dataset' is recorded while anomalies caused by an intruder occur. The test dataset is an unbalanced collection of normal sensor readings (negative class) interleaved by anomalous sensor readings (positive class). Accuracy score is a biased metric since it does not account for the class imbalance. To assess the number of correctly classified instances in the positive class, Recall (True Positive Rate) is an insightful metric that is not biased by class imbalance. 

\subsection{Attacks on Industrial Control Systems}

Related work suggests two approaches to model ICS attacks. \emph{First} as a direct attack to actuators. The attacker is assumed able to gain control over a set of actuators and send malicious control commands that deviate from normal operations. Actuator manipulation can be done stealthily to avoid raising alarms, e.g.,~the attacker manipulates the actuators to slowly cause damages. \emph{Second} as a spoofed signal to sensors as depicted in Figure~\ref{fig:system_architecture}. This type of attack can be modeled in two ways, depending on attackers’ model and intentions. A) In the first case, the signal is spoofed in the control loop. This forces both the controller and the detector to evaluate a wrong system state and consequently, the attacker can masquerade some anomalous operations. B) In the second case, the sensor spoofing occurs in the communication channel between the controller and the anomaly detector. The latter does not have a direct impact on the physical process, but it can be combined with a direct attack on actuators to masquerade anomalies occurring in the system, as demonstrated in~\cite{garcia17hey}.

Among the variety of spoofing attacks that can be designed, in literature this characterization is found: \emph{Replay attack}, \emph{stealthy attack}, and \emph{random attack}. The first class of attacks consists of the spoofing of sensor readings by replaying the historian of values as already occurred into the system. The second class of spoofing is done to diminish the probability of detection by an anomaly detector and comprises: bias injection attacks, where the real value is added/decreased by a Constant bias, and false data injection, where the real value is slowly changed to cause system disruption. Finally, in \emph{random attacks}, the attacker substitutes the value of a sensor reading without engineering of the spoofed value.

Vulnerabilities and attacks to some process-based detectors were demonstrated in the literature. Specifically, in~\cite{erba2020concealment} an in-depth analysis of Deep Learning (model-free) showed how those detectors can be threatened by black-box attackers that can hide anomalies occurring over the system.
In~\cite{dash2019out, quinonezsavior} the resilience of model-based anomaly detectors against sensor spoofing attacks was investigated, specifically they study Control Invariants (CI)~\cite{choi2018detecting} and Extended Kalman Filter (EKF)~\cite{bristeau2010hardware}.
The authors of~\cite{dash2019out} and ~\cite{quinonezsavior} draw different conclusions about EKF resiliency, due to the different threat models considered.

\section{ICS Properties For Anomaly Detection}

\label{sec:properties} 

In this section, we introduce general properties of the physical process sensor data that are often (implicitly or explicitly claimed to be) verified by ICS and in general CPS anomaly detection systems. So far, no consistent understanding of those properties was proposed in prior work. We will introduce those properties of ICS giving an interpretation of their relation with systems in state-space representation. In particular, those properties of the multivariate time series of sensor data are based on the physical process itself, e.g.,~the underlying laws of physics.  To find \up values that can successfully evade an unknown anomaly detector, an attacker needs to exploit those signal properties. 
Finally, we build a taxonomy of process-based anomaly detection from literature w.r.t. the identified properties. A specific system and attacker model is introduced together with our attacks in Section~\ref{sec:system}.

\subsection{Spatial Consistency}
Sensor readings coming from a physical process are correlated among each other (referred to as attribute correlations in~\cite{illiano2015detecting}). Correlation depends on the sensed physical process and control action. 

Considering the System-state representation of a Linear Time Invariant system (see Section~\ref{sec:background}), the output $y_k$ is observed by a set of sensors. Given the state $x_k$ and input $u_k$ of the physical system, the output features are correlated among each other according to the coefficients of $C$ and $D$ matrices.

Spatial consistency of sensor readings can be verified, for example, by explicitly knowing  $C$ and $D$ or monitoring the process cross-correlation function among sensors. An anomaly detector should correctly exploit those correlations to identify anomalies occurring over the system, as anomaly would be generated by a change in $C$ or $D$ coefficients. To succeed in a \up attack against an anomaly detector that accounts for spatial consistency, an attacker should be able to spoof sensor readings in a spatially consistent way (i.e.,~spoofing signal should not break underlying correlations among sensor readings).

The sensor data of a system under attack is often correlated to physical process states that can be observed with sensors out of reach of the attacker. We can explain it with an intuitive example:consider an attacker replaying CCTV recordings of a camera in a public place. The weather and daylight at that location can also be observed with external sensors. Therefore, continuous Replay of prior video data can lead to inconsistencies that can be detected easily. The same holds in an ICS, the same process is measured with multiple sensors correlated, and an attacker replaying the value of just one sensor would cause inconsistencies. 

\subsection{Temporal Consistency}
The temporal evolution of a sensor reading unfolds according to the process physics and control action. 

Considering a System-state representation of a Linear Time Invariant system (see Section ~\ref{sec:background}), the first order derivative of the state $x_{k+1}$ captures the temporal dependence between $x_{k+1}$, $x_k$ and $u_k$ according to $A$ and $B$ matrices. This relation can be observed as the output of the system at time $y_{k+1}$.

Anomaly detectors should check the temporal evolution of the sensed value to verify if a spoofing attack is occurring over the system. From the attacker perspective, spoofed sensor values need to be temporally consistent in order to accomplish their goal. Temporal consistency should be provided not only within the spoofed data but also between the legitimate data and the spoofed data at the start and end of the attack.

For example, in a Replay attack, while data within the recorded time window can be expected to be temporally consistent, the attacker needs to ensure that at the start of the Replay attack, the replayed data is consistent with the previous time step. If this condition does not hold, the Replay can trigger an anomaly detector that monitors sensors temporal consistency.

\subsection{Statistical Properties}

Sensor readings in the analyzed multivariate temporal series are characterized by statistical properties. Those properties are derived from the process that is generating the data, i.e.,~the matrices $A,B,C,D$ of the system in State-space representation.  The process generating the sensor readings can be identified using model identification techniques. Anomaly detection can leverage statistical properties to spot anomalies. For example, each sensor reading is characterized by a proper mean and standard deviation.

Moreover, the measured values are affected by the sensor that does measurement. The sensor itself is characterized by a measurement error that depends on the hardware used. From the unique characteristic of the sensor measurement, fingerprinting the sensors to disclose spoofing attacks has been proposed~\cite{ahmed18noiseprint}. On the other hand, sensors can be prone to drifts. Those drifts cause a change in the statistical properties of the signal. This may cause the raise of false alarms due to outdated statistical parameters.

\subsection{Taxonomy of Process-based Detectors}
\label{sec:taxonomy}
We now present and apply a taxonomy to relevant prior work on process-based anomaly detectors to discuss how that work addressed the challenges outlined above. For each prior work, we briefly summarize how the detection scheme works, report the threat model considered by the defense and analyze which properties among Spatial consistency, Temporal consistency, and Statistical properties are considered as detection features. We also summarize on what kind of system the detector was proposed to operate. In addition, we report if open source code is available for the proposed detectors. Our results allow to identify which properties are (implicitly or explicitly) captured by the anomaly detectors (see Table~\ref{tab:taxonomy}). 

\begin{table}
\setlength{\tabcolsep}{1.4pt}
\begin{center}
\caption{Taxonomy of related works in Process-based anomaly detection. Consistency: \tild partially considered by the detector. Dataset and Process: \TDot simulated process/data, \Tdot real process/data, \Thdot simulated and real process/data
}

\begin{tabular}{r c|c|c|c|c|c|c|c|c}

&    \rota{\cite{aoudi18truth} Aoudi \etal} & \rota{\cite{feng2019systematic} Feng \etal} & \rota{\cite{chen2018learning} Chen \etal} & \rota{\cite{urbina16limiting} Urbina \etal}&\rota{\cite{choi2018detecting} Choi \etal} & \rota{\cite{quinonezsavior} Quinonez \etal} & \rota{\cite{adepu2016distributed} Adepu \etal} & \rota{\cite{ahmed18noiseprint} Ahmed \etal} &\rota{\cite{hau2019exploiting} Hau \etal}\\

\toprule{\textbf{Model-free}} & \cmark & \cmark & \cmark & \cmark&  \xmark    & \xmark  &\xmark & \cmark& \cmark\\
\midrule{\textbf{Consistency}}\\
Spatial                                 & \xmark  & \cmark &   \cmark   & \xmark &   \cmark & \cmark &   \cmark  &   \cmark    & \cmark \\
Temporal                                & \cmark  & \cmark &   \cmark   & \cmark &   \cmark & \cmark &  \tild   &   \cmark    & \cmark \\
Statistical                             & \tild   & \tild  &   \cmark   & \tild  &   \xmark & \tild  &   \xmark &   \cmark    & \tild\\

\midrule{\textbf{Evaluation}}     \\
SWaT                                    & \Tdot & \Tdot & \TDot     &   \Tdot &      -  &  -     & \Tdot & \Tdot &   -  \\
WADI                                    &   -   & \Tdot &     -     &   -     &      -  & -      &   -   & \Tdot &   -    \\
Ten. Eastman                            & \TDot &   -   &     -     &   -     &      -  &  -     &   -   &      -     &   -  \\

Private/Other                           & \Tdot &   -   &     -     & \Thdot  &      -  & -      &   -   &      -     & \Tdot\\
UAV                                     &   -   &   -   &     -     &   -     & \Thdot  & \Thdot &   -   &      -     &   -  \\

\midrule{\textbf{Code Public}}          & \cmark &   \xmark   &  \tild    &   \xmark   &     \cmark  &  \cmark  &   \xmark   &      \xmark     &   \xmark  \\

\bottomrule

\end{tabular}

\label{tab:taxonomy}
\end{center}
\vspace{3pt}

\end{table}

\Par{Aoudi \etal CCS'18~\cite{aoudi18truth}}  In this work the authors introduce PASAD, a model-free anomaly detector. The key motivation of this anomaly detector is to learn and obtain a mathematical representation of the regular dynamics/deterministic behavior of sensor signals in the ICS and spot any anomalous deviation. Singular Spectrum Analysis is used to analyze sensor readings and learn how the process behaves in normal conditions. Sensor readings are considered as a univariate time series. Training is conducted on a set of \textit{lagged} sensor readings, those sensor readings are processed with Singular Value Decomposition (SVD), and projected into a Signal Subspace, where they form a cluster of points. At test time, test samples are processed with the same pipeline and compared against the centroid of the training data cluster in the Signal Subspace. If tested points exceed a distance threshold from the centroid, the system state is evaluated as anomalous. Validation is done on the simulated Tenessee-Eastman Process, SWaT dataset, and Water distribution plant in Sweden. The framework is implemented in MATLAB and is available as open-source. 

\underline{\textit{Threat model.}} The attacker is able to manipulate process data, using system models.  The attacker tries to hide attacks injecting false data (even at the process level) such that the compromised sensor values remain within the noise level.

\underline{\textit{Properties analyzed by the detector.}} \emph{Spatial Consistency:} no, because every sensor is processed individually as a univariate time series processed. \emph{Temporal Consistency:} yes, the signal subspace is computed according to \emph{lagged} timesteps of the sensor readings. This accounts for the evolution of the sensor readings over time. 
\emph{Statistical properties:} partially, the centroid in the signal subspace is calculated projecting the sample mean of the training lagged vectors. This results in projecting the average sensor update into the signal subspace. Implicitly the detector accounts for sensor update mean and embeds the standard deviation within the threshold (diameter of the cluster of normal operation points around the centroid).

\Par{Feng \etal NDSS'19~\cite{feng2019systematic}} This work proposes a framework for automatic process invariants extraction. Those invariants are used to build a model-free anomaly detector. The framework applies Data mining techniques to extract invariants starting from frequent itemset with multiple minimum supports. The framework is divided into two main steps, Predicate Generation and Invariant Mining. During predicate generation, three kinds of predicates are considered. Categorical predicates are generated according to actuators states.  
Distribution-driven predicates, use Gaussian Mixture Models to identify the K probability distributions that describe the sensor value update (this analysis is done per-sensor). Event-driven predicates are fitting a Lasso regression models to capture the interplay between events occurring over actuators states and values seen by sensors. 
Predicates are combined to create itemsets, i.e.,~the set of predicates that hold in a certain instant over the ICS. Invariant Mining extracts frequent itemsets with the CFP-growth++~\cite{kiran2011cfp-growth++} algorithm. 
Evaluation is conducted over WADI and SWaT dataset. No open-source code is available.

\underline{\textit{Threat model.}} The attacker is able to manipulate process data, using system models. In particular, the authors state that the attacker ``[\dots] is considered as an insider and she has the process, communication knowledge, and access to the communication channels. [\dots] In order to achieve her goal, an attacker performs strategic manipulation of sensor measurements and strategic control of actuators''~\cite{feng2019systematic}.

\underline{\textit{Properties analyzed by the detector.}} \emph{Spatial Consistency:} yes, for every timestep the invariant mining algorithm tries to capture inter-dependencies among sensor readings in the given timestep (i.e.,~the invariant). \emph{Temporal Consistency:} yes, temporal evolution of sensor updates is captured in the Distribution Driven predicates. \emph{Statistical Properties:} partially, distribution Driven predicates are computed using Gaussian Mixture Models. This method clusters sensor updates (first derivative), capturing statistically sensor updates distributions.

\Par{Chen \etal S\&P '18~\cite{chen2018learning}} This work applies a different methodology to train a model-free detector, specifically they propose a way to generate anomalous data. First, data of the ICS are generated by running a normal PLC program. Then, a set of mutated PLC programs is generated. Those mutations are used to produce anomalous data. A Support Vector Machine (SVM) is trained to distinguish the two classes. Data of the time series are presented to the Support Vector Machine in couples of ($\pi$, $\pi'$) where $\pi'$ is the measured sensor readings after $d$ time-steps. A trace ($\pi$, $\pi'$) is anomalous when it is classified as such by the anomaly detector. The evaluation was done on SWaT simulator; the code of simulator and code mutation is available online, anomaly detector and data are not.

\underline{\textit{Threat model.}} The attacker is able to manipulate process data, using system models. Two kinds of attacker models are taken into account in the contribution. Network attacks, in which an attacker can manipulate network packets containing sensor readings and actuator commands. Code-modification attacks, in which the attacker can randomly modify the different PLC programs in the simulator.

\underline{\textit{Properties analyzed by the detector.}} \emph{Spatial Consistency:} yes, the SVM is trained on multivariate sensor vectors. \emph{Temporal Consistency:} yes, the SVM is trained on couples of sensor readings collected at different time steps, to capture the temporal evolution.  \emph{Statistical Properties:} yes, SVM does capture the statistical properties of the traces ($\pi$, $\pi'$) implicitly.

\Par{Urbina \etal CCS '16~\cite{urbina16limiting}} In this work, a method to limit Stealthy attacks is presented together with a detection metric to compare anomaly detectors. The detection metric accounts for the expected time between false alarms and maximum deviation imposed by undetected attacks per time unit.  Model-free Auto-Regressive (AR) and model-based Linear Dynamical State-space (LDS) methods are applied to evaluate the proposed metrics. AR models are used to predict the output of the system, given the last N sensor measurements. LDS systems are used to predict the output of the system starting from control commands and system state. Stateless tests (e.g.,~such as a comparison with a threshold) and stateful tests (e.g.,~such as Cumulative Sum (CUSUM)) are compared as detection statistics. Results show that stateful statistics provide a better detection performance w.r.t. to those metrics. 

\underline{\textit{Threat model}} The attacker is able to manipulate a specific process feature, using system models. In particular, the authors state that the attacker ``has compromised a sensor or an actuator [\dots] The adversary [\dots] knows the physical model we use, the statistical test we use, and the thresholds we select to raise alerts. Given this knowledge, she generates a stealthy attack, where the detection statistic will always remain below the selected threshold''~\cite{urbina16limiting}.

\underline{\textit{Properties analyzed by the detector.}} \emph{Spatial Consistency:} Partially, the AR model considers univariate time series while the LDS model captures the correlations between the sensor readings.  \emph{Temporal Consistency:} yes, prediction is done one step ahead. Prediction error would then trigger an alarm in Stateful and Stateless detection statistics.  \emph{Statistical Properties:} partially, statistical properties of the are captured by the prediction error score in Stateful tests.

\Par{Choi \etal CCS '18~\cite{choi2018detecting}} This work proposes a model-based detector based on PID controllers to approximate Robotic Vehicle's (RV) control logic. The PID controller predicts the system state, if the the prediction error surpasses a threshold an alarm is raised. The error is monitored in time windows of length $w$. $w$ is tuned to overcome the latency that can be introduced by the PID. 
Implementation is available as online.

\underline{\textit{Threat model.}} The attacker manipulate  process data, in the words of the authors ``interfere with RV operations by corrupting or injecting (actuation or sensor) signals through external means (e.g.,~distorting actuation signals or misleading sensors to generate erroneous readings)''~\cite{choi2018detecting}.

\underline{\textit{Properties analyzed by the detector.}} \emph{Spatial Consistency:} yes, the PID controller accounts for system state. The system state comprises various sensor readings. \emph{Temporal Consistency:} yes, the prediction is computed one step ahead, deviations in the temporal evolution would be captured when the prediction error is computed. \emph{Statistical Properties:} no, the PID model does not account for the statistical properties of the signal.

\Par{Quinonez \etal USENIX Security '20~\cite{quinonezsavior}} In this work the use of Extended Kalman Filters (EKF) is proposed to detect anomalies in Autonomous Vehicles in a model-based manner. Specifically, the system exploits the knowledge of the physical dynamics of the system to extract physical invariants. Those invariants are learned in an offline manner, collecting data from the vehicle in motion. The prediction of the EKF, combined with CUSUM change detection algorithm allows the detection of anomalies.
Implementation is available online. 

\underline{\textit{Threat model.}} The attacker is assumed to be able to inject false signals to sensors and actuators. Attacks that inject false signals in all the sensor and actuators are considered out of scope. The detection mechanism requires at least one sensor/actuator couple that reports true data to the detector.

\underline{\textit{Properties analyzed by the detector.}} \emph{Spatial Consistency:} yes, the EKF models the spatial correlations among sensors and actuators. \emph{Temporal Consistency:} yes, the prediction is computed one step ahead. Moreover CUSUM algorithm computes a windowed score. \emph{Statistical Properties:} implicitly, EKF models the expected dynamics of the physical model, hence its statistical properties.

\Par{Ahmed \etal AsiaCCS '18~\cite{ahmed18noiseprint}}
In this work, the authors tackle the problem of intrusion detection from the fingerprinting perspective. The main idea is to detect anomalies occurring in the system from changes in the noise in sensor readings. The paper proposes a mix of model-based and model-free techniques. To obtain a sensor fingerprint, Kalman filtering is used. 
The obtained prediction is compared with the real state to produce residuals. Those residuals are used to produce features to train a Support Vector Machine classifier. 
Implementation is not available as open-source.

\underline{\textit{Threat model.}} The  attacker  is  able  to  manipulate  process data,  using  system  models.  In  particular,  the  authors  state that ``the attacker knows the system dynamics,
the state space matrices, the control inputs and outputs, and the
implemented detection procedure. 
[\dots] An attacker
can compromise these communication links in a classic Man-in-The-Middle (MiTM) attack. [\dots] For data injection attacks, it is considered that an attacker injects or modifies the real sensor measurement''~\cite{ahmed18noiseprint}.

\underline{\textit{Properties analyzed by the detector.}} \emph{Spatial Consistency:} yes, the system model captures the spatial consistency to predict the system state. \emph{Temporal Consistency:} yes, Kalman filter predicts system state one step ahead. \emph{Statistical Properties:} yes, Support Vector Machine is trained with the statistical properties of the signal and prediction residuals.

\Par{Adepu \etal AsiaCCS '16~\cite{adepu2016distributed}}
In this work, the authors present a detection scheme based on the usage of process invariants. The study focuses on the so-called `State-dependent' and `State-agnostic' invariants. 
Experiments are conducted in the SWaT testbed. Code is not available as open-source.

\underline{\textit{Threat model.}} The attacker can perform denial-of-service attacks and stealthy attacks. Specifically the attacker can tamper with sensor and actuators and manipulate the information exchanged within the network.

\underline{\textit{Properties analyzed by the detector.}} \emph{Spatial Consistency:} yes, `State-dependent' invariants capture the spatial correlations among sensor readings at a certain time step.  \emph{Temporal Consistency:} yes, `State-agnostic' invariants for water tanks are modeled as one step ahead prediction based on the previous state, inflow, and outflow of the tank. \emph{Statistical Properties:} no, the detector does not account for signal statistical properties.

\Par{Hau \etal CPSS '19~\cite{hau2019exploiting}}
In this work, the authors consider the problem of false data injections in low-density wireless sensor networks. They develop a model-free anomaly detector that is exploiting temporal-attribute correlation among sensor readings. They extract pairwise feature cross-correlation and check if this correlation is statistically supported by the stream of data. When cross-correlation indices fail the test, an alarm is raised. 

\underline{\textit{Threat model.}} The  attacker  is  able  to  manipulate  process data, and can have two possible goals. ``The goal of the attacker is to cause an undesired system response by spoofing a sensed system condition that the system would react to. [...] The goal of the attacker is to hide an undesirable system condition to prevent a system response''~\cite{hau2019exploiting}.

\underline{\textit{Properties analyzed by the detector.}} \emph{Spatial Consistency:} yes, cross-correlation function spatially relates sensor readings. \emph{Temporal Consistency:} yes, time series of cross-correlations are analyzed to detect anomalies in the time evolution of the signal. \emph{Statistical Properties:} partially, statistical properties of the signal are implicitly evaluated by the method.

\subsection{Summary on Properties Verified for Anomaly Detectors}

As shown in our taxonomy, every reviewed anomaly detector applies some techniques that, implicitly or explicitly, capture at least one among Spatial consistency, Temporal consistency, and Statistical properties. The provided analysis confirms that the three identified challenges represent the common set of properties that anomaly detectors need to correctly model to detect anomalies occurring over the physical process. 

\section{\up Attacks }
In this section we present our \emph{\up} attacks. We call them Synthetic because they are built without explicit knowledge of the physical model of the system and can be pre-computed offline by the attacker. We introduce our considered System and Attacker Model that fit prior work assumptions, explain our problem statement, provide a motivating example and present our abstract framework design. Finally we position our \up with prior work stealthy attacks.
\label{sec:system}
\subsection{System Model}

We consider an Industrial Control System as depicted in Figure~\ref{fig:system_architecture}, consisting of a number of sensors and actuators, connected to a controller (e.g.,~PLC). The controller reports local sensor data to an anomaly detector that uses the sensor data to classify the system state as anomalous or normal.

\subsection{Attacker Model}
We assume an  attacker aiming to hide an anomaly in the physical process by \up (e.g.,~using a long-term Man-in-the-PLC attack~\cite{garcia17hey}) without knowing the physical properties of the system (i.e., lower requirements than prior work attackers). The attacker needs to evade a process-based detection system (in contrast to~\cite{garcia17hey} where a human operator is considered). 
Instead of a noisy brute force attack, the attacker aims to perform a stealthy attack (such as Stuxnet~\cite{weinbergerStuxnet}) to increase the overall damage caused over time, and cover attacker's tracks. The actual anomaly produced by the attacker is out of the scope of this paper and could have been introduced by the attacker over the network (e.g.,~through injection of malicious actuator commands), caused by the attacker through other physical channels (e.g.,~physical interaction), or could have occurred naturally~\cite{wadi17dataset, mathur16swat, taormina18battle}. 

As in the threat models summarized in Section~\ref{sec:properties} the attacker is assumed to deterministically manipulate (a subset or all) sensor readings processed by the anomaly detector (e.g.,~through traffic manipulation~\cite{urbina16fieldbus}, or by remote sensor spoofing~\cite{kune2013ghost,yan2016can,tu19trick}). The attacker has also the ability to eavesdrop the values that are transmitted from sensors or measure the process directly. We highlight that our attacker model has lower requirements than attacker models proposed in prior work and summarized in Section~\ref{sec:properties}. Our attacker does not know the physics of the process or the thresholds used by the anomaly detector. Since the detection mechanism requires data aggregation and transformation, the natural choice for anomaly detection deployment is the historian server (in an ICS). As practically demonstrated in~\cite{garcia17hey} PLC data transmission to the historian offers the attack surface that an attacker can leverage to spoof sensor data and conceal the true state of the system.

\subsection{Problem Statement}
Our goal is to evaluate the resiliency of process-based model-free detectors when targeted by \up attacks. In particular, we aim at probing whether the anomaly detector leverages physical properties of the system to detect anomalies. We assume an attacker that launches their attacks over the physical process to cause an anomaly. Simultaneously, to avoid detection, the attacker spoofs sensor readings seen by the process-based detector. To assess the resiliency of detectors, the attacker generates spoofed values according to different strategies. In particular, we are interested in understanding the properties of the spoofed signal that evades detection. Moreover, as a second objective, the attacker aims to minimize the number of spoofed sensor readings. 

\subsection{Motivating Example}
We now present a motivating example scenario based on our attacker and system model. In a distributed Power Grid ICS, an attacker gains physical access to a remote substation of the distribution system and installs a Man-in-the-PLC~\cite{garcia17hey} malware in the PLC. Using that malware, the attacker can inject commands to actuators, and send spoofed data to the SCADA (as most industrial protocols do not feature security mechanisms such as authentication or confidentiality). The SCADA uses an anomaly detector to detect manipulations of the physical process, such as the ones proposed in~\cite{aoudi18truth,feng2019systematic}. The attacker knows that a process-based detection system is used, and starts the attack by passively eavesdropping process data to learn normal operations. Then, the attacker starts to send malicious actuation commands (to damage the process, out of the scope of this paper), and manipulates the sensor data reported to the SCADA to hide the ongoing attack from the process-based anomaly detector. 



\subsection{Abstract Framework Design}

\begin{figure}
    \centering
    \includegraphics[scale=0.8]{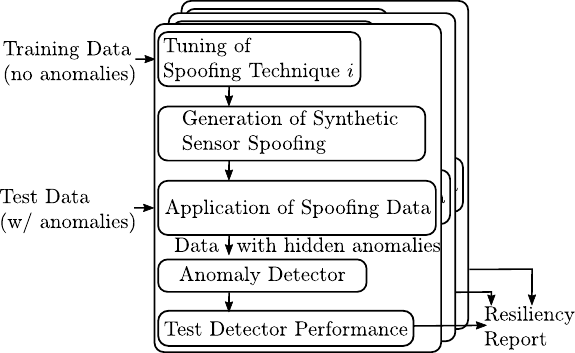}
    \caption{Overview of usage of Framework for \up for resiliency analysis of an anomaly detector. The framework receives as input eavesdropped data (containing no anomalies) and test data (containing anomalies). It then applies \up techniques to the data. Spoofed data are evaluated with the anomaly detector. As a result, the performance of the detector is calculated in the resiliency report.}
    \label{fig:resiliency_analysis}
\end{figure}

Our approach tests the anomaly detector in a black-box fashion to assess its resiliency properties against \up attacks as depicted in Figure~\ref{fig:resiliency_analysis}.  Given a target anomaly detector, our analysis requires eavesdropped data and anomalous data.  
We then tune the \up technique based on the eavesdropped data, generate \up examples, and apply them to the (anomalous) test data.
To perform the evaluation, we assess the performance of the detector over data containing anomalies (test data). Then, we test the different \up attacks. If an attack succeeds, the performance of the detector decreases and the detector is observed vulnerable to that \up technique. \up attacks are designed not only to attack detectors and reduce their Recall score but also to highlight which properties the detector fails to capture properly.

\subsection{Other Physics-Aware Stealthy Attacks}
Our attacks are able to evade model-free process-based detectors i) without knowledge of the physical process and its underlying physics, and ii) can be precomputed offline by the attacker. This assumes the same attacker capability as in the Man-in-the-PLC attack~\cite{garcia17hey}, where the attacker spoofs sensor readings received by the SCADA system from the human operator. In contrast to our work, \cite{garcia17hey} assumes an adversary that needs to  know the inner workings of their target, exploiting this information they build a model of the correct behavior of the target to hide effects of the attack to the physical process.

Other prior work attacks cannot be precomputed, and have to be dynamically optimized at runtime to stay below the currently estimated detection threshold~\cite{urbina16limiting, quinonezsavior, dash2019out}. As last class of comparable attacks, a number of attacks are only temporally stealthy~\cite{wadi17dataset, mathur16swat, taormina18battle} until the goal is reached. In contrast to those attacks, our attacks allow to hide the attack even after the main attack goal was achieved.

\section{Framework for ICS Spoofing}
\label{sec:framework}
In this section, we give an overview of our \up framework design and implementation. The framework uses three different approaches to generate \up examples: Constant spoofing, Gaussian spoofing, and selective Replay. For each method, our framework supplies a tuning mechanism, a block to generate \up examples, and a method to apply the \up examples to anomalous data. The resulting data with hidden anomalies are tested against ICS anomaly detectors (see Section~\ref{sec:evaluation}).

\begin{table}[tb]
\caption{Spoofing methods in our framework and their expected efficacy against the different consistency checks in prior work. \Tdot expected to be efficient,
\Thdot expected to be efficient for weaker constraints.}
\begin{center}

\begin{tabular}{r ccc} \toprule
                                 &\multicolumn{3}{c}{Expected to Spoof}     \\ 
Method                           &    Spatial    & Temporal&  Statistical     \\ \cmidrule(l){1-1} \cmidrule(l){2-4}
Constant                         &    -        & \Tdot &   \Thdot     \\
Gaussian                         &    -        & -     &   \Tdot      \\
Replay                           &    \Tdot    & \Thdot &  \Tdot      \\

\bottomrule
\end{tabular}

\label{tab:methods}
\end{center}
\end{table}

\subsection{Constant Spoofing}
The most basic form of \up is based on Constant spoofing of values for each attacked sensor. Tuning of those \up examples only requires the computation of the average value of the sensor over a long-time window (i.e.,~the whole training data). While this approach sounds very simple, we show later in Section~\ref{sec:evaluation} that the tested detectors seem to be designed to tolerate such Constant values (e.g.,~as they often occur in normal operations). As the sensor values are Constant, temporal consistency can be expected to hold (e.g.,~no sudden jumps in the sensor values). Each sensor value is spoofed individually, so spatial consistency is not necessarily preserved. Nevertheless, if all correlated values are spoofed, the relation between the constant average values could potentially meet the expectations of the detection. Clearly, the spoofed signal will have an easy-to-recognize distribution, so statistical consistency analysis might detect such spoofing. As we show in Section~\ref{sec:evaluation}, statistical consistency checks that only check average or ranges of sensor values can be expected to fail to detect such spoofing. Table~\ref{tab:methods} summarizes the expected efficacy of our spoofing methods w.r.t. the consistencies defined in  Section~\ref{sec:properties}.

\subsection{Gaussian Spoofing}
Our Gaussian spoofing attack changes the real values seen by the detector by generating sensor values that mimic the statistical distribution of the original sensor signal. For the tuning of these \up examples, we compute the average $\mu$ and standard deviation $\sigma$ for each sensor value (over the whole training data). Then, our spoofing signal consists of samples drawn from a Gaussian distribution with parameters ($\mu, \sigma$). We expect Gaussian spoofing to not be temporally consistent, as from one sample to the next the value of the sensor might jump (in contrast to the expected behavior of low noise sensor readings). As with the Constant spoofing, spatial consistency is also likely violated. As our spoofing signal closely matches the statistical properties of the original signals, we expect the spoofing to be statistically consistent.

\subsection{Replay Spoofing}
This block implements a variant of Replay Spoofing (e.g.,~discussed in~\cite{moSinopoli2009}). The attack spoofs sensor values according to sensor readings observed in the past (e.g.,~by signal eavesdropping). We expect Replay spoofing to be Spatially and Statistically consistent as it replicates normal operations occurring over the system. If the attacker is constrained to spoof only a subset of sensors, then Spatial Consistency might also be violated. Regarding Temporal consistency, it is likely to be violated at the beginning and at the end of the Replay, as the replayed values are not consistent with the real state of the system at the beginning and end of the spoofing.

\subsection{Implementation}
We implemented the \up framework using Python 3.7.3, using the Pandas and NumPy libraries. The framework processes as input the training data (without anomalies) and test data (containing anomalies). Data should be organized in .csv format, where every row contains the sensor readings collected at a certain time step and every column represents a different sensor value. Test data is labeled, indicating whether the given row was anomalous or not. Optionally (in the constrained case) the framework takes the constraints on which variables can be spoofed. It then applies the presented \up techniques to the data. 

Starting from the test data, we first identify the intervals in the dataset that contain the attacks. The \verb|spoofing| function builds the dataset containing \up. It leaves unchanged the time intervals where ground truth reports `normal' and applies the spoofing to the time steps with ground truth `anomalous'. The different \up techniques are implemented as functions that apply the required spoofing to the given data. \verb|Constant| spoofing, for every sensor, computes the mean or median of the sensor reading in the eavesdropped data and substitutes it to the data. \verb|Gaussian| computes mean $\mu$ and standard deviation $\sigma$ and the substitute data with samples from $\mathcal{N}(\mu, \sigma)$. \verb|Replay| copies the data as found in the eavesdropped dataset. We plan to release our framework implementation available as open-source.

\section{Experimental Evaluation Setup}
\label{sec:evaluation}
In this section, we explain how we used our framework for the resiliency analysis of \num anomaly detectors. To assess the resiliency of an anomaly detector we perform a resiliency analysis as depicted in Figure~\ref{fig:resiliency_analysis}.  The resiliency analysis takes as input the following elements: eavesdropped data (benign data that the attacker has recorded from the ICS traffic), test data (data containing anomalous sensor readings that the attacker aims to falsify to avoid detection), the anomaly detector under test. Resiliency analysis applies different \up techniques to the test data and assesses the performance of the anomaly detector over spoofed data. Once all the different techniques were tested a results report is generated.

For every designed \up technique, we evaluated the system under attack in two different settings whether possible. The first, unconstrained attack, in which the attacker can spoof all the sensor readings to deceive anomaly detection. The second, constrained attack, in which the attacker can spoof only a limited number of sensor readings. 

We evaluated our \up techniques over \num model-free anomaly detectors for ICS recently proposed at security conferences and introduced in Section~\ref{sec:taxonomy}.  The first~\cite{aoudi18truth} is available open-source and implemented for MATLAB. The second~\cite{feng2019systematic} was not available as open-source, and re-implemented it for this paper. The third~\cite{urbina16limiting} was not available as open source and we re-implemented it. The fourth~\cite{chen2018learning} was provided to us by the authors of the paper together with the data used in their contribution. We plan to release our re-implementations as open-source. We note that~\cite{quinonezsavior, choi2018detecting} are model-based detectors and hence out of scope for our analysis. 

\subsection{SWaT Dataset}
Secure Water Treatment (SWaT)~\cite{mathur16swat} is a room sized ICS testbed built for Cybersecurity research at Singapore University of Technology and Design. Water treatment comprises six-stages, raw water is treated to obtain drinkable water. The stages comprise: chemical dosing, ultrafiltration, de-chlorination and reverse osmosis. Water treatment is controlled by cyber components that comprise sensors and actuators that communicate to PLCs where the control logic resides, PLCs are interconnected together in a ring network. Every PLC is then connected to the SCADA system where the system state is displayed through HMI and stored in the Historian. 

SWaT dataset~\cite{goh2016dataset} contains the data collected from testbed sensors and actuators during 11 days of continuing operations. Specifically, 51 attributes are collected, 24 continuous sensors and 27 integer actuator states with sampling time 1 second. During the first 7, the system operates in normal operating conditions while in the remaining 4 days 36 attacks were launched. Every attack comprises different physical manipulations of the system, both for localization and purpose.

\subsection{BATADAL Dataset}
BATADAL Dataset was released as part of the BATADAL competition~\cite{taormina18battle}. The dataset is generated through EpanetCPA~\cite{taormina2019epanetCPA}, an open-source MATLAB toolbox for water distribution attack simulation. The water distribution network simulated in the dataset is C-town~\cite{ostfeld12ctown}. Data are divided into three sets; the first containing 1 year of simulation under normal operating conditions, the second and the third containing 14 attacks (7 attacks each). 43 variables are captured by the dataset, continuous values for sensor readings and discrete for actuators with sampling time 1 hour.
An updated version of the dataset was released in~\cite{erba2020concealment} and available as open-source at \url{https://github.com/scy-phy/ICS-Evasion-Attacks}. In the updated version, the sampling time was reduced to 15 minutes, and in the attack data, no spoofing techniques are applied (the original dataset had some spoofing applied). For our experiments, we refer to this updated version of the dataset, since our goal is to conceal existing anomalies in the dataset with our \up attacks.

\subsection{Implementation of Aoudi et al.}
The implementation of PASAD anomaly detector~\cite{aoudi18truth} is available as open-source at~\url{https://github.com/mikeliturbe/pasad}. PASAD analyzes every sensor univariate temporal series independently, for every sensor PASAD requires to be trained independently. We trained PASAD on the SWaT dataset LIT301 sensor, using the parameters indicated on the original paper. Specifically, we used $N = 30000$, $L = 5000$, $r = 10$.  Since no exact value for the threshold was given by the authors, we deduced the threshold from form their result graph as $\theta = 3\times10^6$. The resulting detection performance with this threshold over sensor LIT301 is  Accuracy = 0.88, 
F1-score = 0.55, 
Precision = 0.49, 
Recall(TPR) = 0.63
, FPR = 0.09. 

\subsection{Implementation of Feng et al.} \label{sec:implementation} 
As the detector of~\cite{feng2019systematic} was not available as open-source (or on request), we had to re-implement the anomaly detector based on the paper. We used Python 3 with the following libraries: Sklearn, Pandas, NumPy, SciPy. In this section, we summarize the parameters, and the assumptions we had to make to implement the detection system. We note that we contacted the authors of~\cite{feng2019systematic},  but did not get details on all parameters used in the original paper. \Par{Distribution Driven Strategy}
We normalized the data between 0 and 1. For every sensor, we fitted Gaussian Mixture Models with at most 4 components and took the one with the lowest BIC score. \Par{Event Driven Strategy} We set the threshold for the trigger $\epsilon$ = $0.05$, for Lasso we set $\alpha$ = $0.1$. \Par{Invariant Mining}
Invariant mining is done with the CFP-growth++ algorithm. This algorithm is only available as open-source in a Java library\footnote{SPMF \url{http://www.philippe-fournier-viger.com/spmf/}}. We used that library from our python script. Since the library is generating all the frequent itemsets that have the allowed minimum support, we parsed the output to identify the itemsets that do not break the non-redundant condition.

After implementing the detection mechanism, we were unable to reproduce the results of~\cite{feng2019systematic} over SWaT and WADI datasets. Nevertheless, we were able to achieve a comparable result using the BATADAL dataset. The results are as follows. Accuracy = 0.93 
F1-score= 0.58 
, Precision = 0.75
, Recall (TPR) = 0.47
, FPR = 0.02.

\subsection{Implementation of Urbina et al.}
As the detector and data of~\cite{urbina16limiting} were not available as open-source (or on request), we re-implemented the model-free detector with stateful test presented in the paper. The detector is based on an Auto Regressive (AR) model trained over an univariate time series. The residuals of the AR model are used to compute a Cumulative Sum (CUSUM) statistic whose objective is to reveal a change in the process generating the data, i.e.,~spot anomalies in the system. We implemented this detector in MATLAB, using the System Identification Toolbox. We trained the model over BATADAL data, we performed our experiments on sensor `PRESSURE J302', i.e.,~the sensor that (alone) was disclosing the highest number of attacks (8 attacks over 14) with the considered detection method. We selected the AR model of order 20 with Best Fit criteria and tuned the CUSUM parameters using grid search and selected \verb|control limit| = 5.5 and \verb|min mean shift detect| = 1 obtaining Accuracy = 0.91, 
F1-score = 0.41, 
Precision = 0.79, 
Recall = 0.28, 
FPR = 0.01. 

\subsection{Implementation of Chen et al.}
The detector is not available as open source. The authors of the paper kindly provided us the SVM detector and the data used in their contribution. The data are collected from a simulator of the SWaT testbed, and comprise $3$ sensor values. Differently for other datasets from prior work, the train dataset is composed by both normal and abnormal traces as generated by the mutated PLC programs. 
The SVM model is trained on traces ($\pi$, $\pi'$) where $\pi'$ are the sensors delayed by $100$ time steps, resulting in a $6$ features vector. The performance of the provided detector is the following Accuracy = 0.98 , F1-score = 0.98, Precision= 0.96, Recall= 1.00, FPR = 0.04.

\section{Evaluation Results}
\label{sec:results}
In this section, we show the results of our experiments. We test our framework against \num model-free anomaly detectors from prior work, specifically Aoudi et al.~\cite{aoudi18truth}, Feng et al.~\cite{feng2019systematic}, Urbina et al.~\cite{urbina16limiting}, Chen et al~\cite{chen2018learning}. Our evaluation is based on the analysis of Recall score before and after the \up attacks. The lower Recall, the more the attacker is stealthy. 

Note on False Positive Rate fluctuation in results: anomaly detectors classify instant $t$ aggregates all the $t-n$ sensor reading occurred before $t$. Datasets are composed of attacks interleaved by normal operations, if we spoof from instant $a$ to instant $b$ the classification outcome at time $b+1$ depends on the manipulation occurred between $a,b$, influencing the FPR. More details about our evaluation methodology can be found in Appendix~\ref{sec:methodology}.

\subsection{Aoudi et al.}

\begin{table}
    \centering
        \caption{Resiliency analysis, Aoudi et al. SWaT Dataset sensor LIT301 (which was used as example in~\cite{aoudi18truth}). Threshold $3\times10^6$. Our Constant and Gaussian attacks reduce the recall from 0.63 to 0.061, without increasing the FPR.}
\begin{tabular}{rccccc}

    \toprule
    Dataset & Acc. & F1 & Prec. & Rec. & FPR \\
    \midrule
        Original 
         & 0.88 & 0.55 & 0.49 & 0.63 & 0.09 \\ 
        Replay  
        & 0.79 & 0.08 & 0.08 & \textbf{0.07} & 0.11 \\ 
        Constant 
        & 0.83 & 0.08 & 0.11 & \textbf{0.06} & 0.07 \\ 
        Gaussian 
        & 0.83 & 0.08 & 0.11 & \textbf{0.06} & 0.07 \\ 
        \bottomrule
    \end{tabular}
    
    \label{tab:results_resiliency_swat}
\end{table}
Starting from the open-source implementation of PASAD anomaly detector trained over SWaT dataset (sensor LIT301), we attack it with our \up attacks. In~\cite{aoudi18truth}, no threshold to assess detection performance metrics was proposed explicitly, a qualitative evaluation from graphical plots was proposed. We start with the same approach, and provide a quantitative analysis afterward\footnote{We tested sensor LIT301 as done in the original contribution. The authors tested the anomaly detector also over AIT202, but the performance was not originally good due to process drift.}.

\Par{Qualitative Evaluation} Due to space constraints our detailed results and plots are shown in Appendix~\ref{sec:qualitative}. Using this visual method, we found that all the proposed \up attacks evade PASAD detection. The only attacks that were not completely hidden by our attacks are the ones happening at the beginning of the data capture (December 28th), and after January 1st. This likely is the case as the system state is left in an anomalous state after the attack ends. To compensate for this, we can continue the spoofing until the system recovers from the attack and returns to a steady state.

\Par{Quantitative Evaluation} To measure analytically the evasion achieved by every attack, we evaluated the performance w.r.t. the threshold deducted from the graphics in the original paper. Table~\ref{tab:results_resiliency_swat} summarizes the performance of PASAD when targeted with our \up. Each proposed \up reduces the detector's performance. Since the dataset is highly unbalanced towards the negative class, we focus on Recall and Precision scores. In all the three proposed \up attacks we note that Recall decreases to 0.07 (or less, from 0.63). Furthermore, we can highlight that our attacks do not increase the False Positive Rate (FPR), so they do not induce False Positives (instead, Constant and Gaussian Noise attacks decrease it).

The results of our analysis done over this anomaly detector show that (although the original results were promising), the anomaly detector is not able to extract physical dynamics of the system. Indeed, the evaluation score decreases when we spoof the signal with a not physically plausible temporal evolution of the signal as in the Gaussian and Constant spoofing. This shows that the anomaly detector has learned the data distribution and not the physical process dynamics.  

Mathematically this phenomenon can be explained by analyzing Step 3 of PASAD anomaly detection scheme. PASAD projects training points in the signal subspace. Those projected points create a cluster in the projection subspace. Then, PASAD tracks the distance from the centroid of the cluster to identify anomalies. The centroid is defined as the sample mean of the lagged vectors. Our Gaussian and Constant spoofing fulfill the requirements of being projected within the cluster in the signal subspace. Despite their dynamics is not plausible, their departure score is lower than the threshold.

\subsection{Feng et al.}
\begin{table}[tb]
\setlength{\tabcolsep}{5.5pt}
    \centering
\caption{Resiliency analysis, Feng et al. Batadal test dataset 1, unconstrained attack}
\begin{tabular}{rccccc}
     \toprule
    Dataset & Acc. & F1 & Prec. & Rec. & FPR \\
    \midrule
        Original 
          & 0.9263 & 0.5760 & 0.7462 & 0.4690 & 0.0191 \\ 
        Replay   
            & 0.8778 & 0.0226 & 0.0774 & \textbf{0.0132} & 0.0188 \\ 
        Constant 
            & 0.8765 & 0.0009 & 0.0032 & \textbf{0.0005} & 0.0187 \\ 
        G.N. 
            & 0.9293 & 0.5983 & 0.7604 & 0.4931 & 0.0186 \\ 
        G.N. v2 
            & 0.8816 & 0.0824 & 0.2390 & \textbf{0.0498} & 0.0190 \\ 
        \bottomrule
    \end{tabular}
    
    \label{tab:results_resiliency_batadal}

\end{table}

\begin{figure}[tb]
    \centering
    \includegraphics[width=\linewidth]{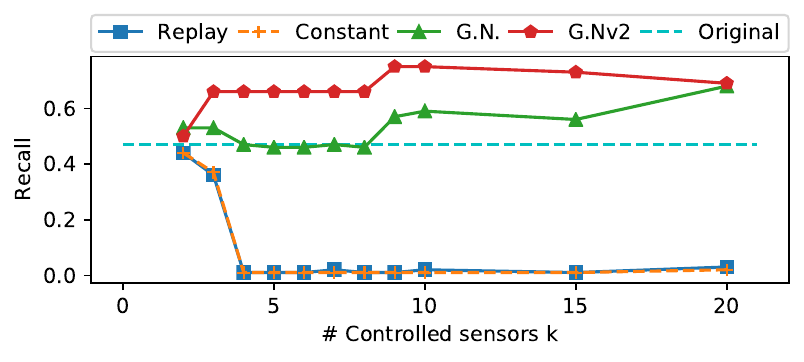}
    \caption{Resiliency analysis, Feng et al. Batadal test dataset 1, constrained attack. This plot shows how the recall score decreases when constrained \up is performed.}
    \label{fig:plot_feng}
\end{figure}

We tested our \up attacks against Feng et al. anomaly detector as described in Section~\ref{sec:implementation}. Since this anomaly detector generates invariants aggregating all the sensor readings, we tested our framework in constrained and unconstrained settings.

\Par{Unconstrained setting} Table~\ref{tab:results_resiliency_batadal} reports the results of unconstrained attack (i.e.,~the attacker can spoof all the sensor values in the network). Starting from the original detection recall of $0.47$, results show that the \up attacks decrease dramatically the Recall score. Applying the Replay attack, the Recall decreases to $0.013$ (expected from prior work). This happens because the spoofed signal is matching normal operations, evading detection. When we apply the Constant attack, Recall drops to $0.0005$. This result indicates that the \up attack was able to conceal the instances of anomalous data. If we consider what is going on in the anomaly detector, Recall close to $0$ means that there are no invariant rules violated by the attack. Indeed, Distribution Driven predicates are not violated since during training it often occurs that a sensor reading remains Constant within two instants. At the same time, Event-Driven Predicates cannot be triggered. The spoofed signal reports that no events are occurring over the system and Spatial correlations are not checked anymore. Hence, the system appears arrested, with no dynamics occurring, and invariant based detector stops checking invariant rules. Again, this demonstrates that the expected dynamic of the system is not captured by the anomaly detector. When we apply the Gaussian Noise attack, results show that the Recall score increases. Looking at the triggered rules in the anomaly detector, we noticed that the actuator states triggered Categorical and Event-Driven invariants. To evade this trigger, we tested Gaussian Noise 2 attack where the spoofing of actuators reports either unknown states (e.g.,~non-integer values for actuators or integers out of the observed range) or Constant values. This aspect highlights an interesting property of Invariant Based anomaly detectors: unknown states do not break any invariant rule as there are no rules that match the unknown prefix bringing The Recall score to $0.049$.

\Par{Constrained attack} The second experiment studied the constrained case where the attacker can only spoof a constrained set of sensor readings, i.e.,~they have compromised a subset of the PLCs and can spoof only certain sensors. The anomaly detector accounts for Spatial consistency through the correlations found in Event-driven predicates and by the frequent itemset mining step. Spoofing of a constrained set of readings would introduce contextual anomalies that should be detected. As depicted in Figure~\ref{fig:plot_feng}, this detector fails to spot the Spatial inconsistencies introduced by Replay and Constant constrained spoofing attacks. Indeed, when the attacker gains control of 4 out of 43 sensors (coming from at most 3 different PLCs out of 9 in the network), the detection recall drops to $0.0137$ in both the cases. Attacks with Gaussian Noise fail to succeed in this constrained setting. Hence, in the constrained case, the detector is resilient to them, as Gaussian Noise introduces Temporal inconsistencies that are detected by invariants. 

\subsection{Urbina et al.}

We attack the autoregressive (AR) detector with our \up attacks. An AR predictor is defined good if it generates residuals (i.e.,~prediction errors) distributed as white noise~\cite{brockwell1991time}. This anomaly detector uses the CUSUM algorithm to check whether the residuals are changing their distribution w.r.t. training phase. When an anomaly occurs, the CUSUM detects a change in the distribution of the residuals (i.e.,~no more distributed as white noise (WN): this is caused by the predictor that is not behaving optimally because of anomalous data). To circumvent this defense, the attacker has to modify the sensor signal in such a way that the obtained residuals do not surpass CUSUM control thresholds, i.e.,~residuals remain WN.

In Table~\ref{tab:results_urbina} we report the results of our \up attacks (results can be visualized in Appendix Figure\ref{fig:urbina_result}). If we consider the Recall rate, we can notice that Replay and Constant attack reduce it respectively to $0.0046$ and $0$. Both approaches are hiding the anomalies from the detector. Replay attack hides anomalies because the replied data are samples of the legitimate temporal evolution of the process. The Constant attack changes the value of the sensor with the average of the process; this forces the AR(20) model to predict the Constant value that is close to the average that sends to $0$ the CUSUM upper and lowers statistics. In the case of the Gaussian Noise attack with $\mu$ = sensor mean, $\sigma$ = sensor standard deviation we found that the spoofed signal is detected, causing an alarm rate higher than in the original case. In this case, the spoofed signal is evolving with no temporal correlation that causes the AR(20) to overshoot the prediction that changes the distribution of the residuals observed by CUSUM. Reducing the standard deviation of the generated Gaussian Noise to a fraction of the original standard deviation produces a signal that evades detection. We tested the following fraction of the original standard deviation of the signal: $0.9\sigma{}, 0.8\sigma{}, 0.7\sigma{}. 0.6\sigma{}, 0.5\sigma{}$. Spoofing the sensor signal with Gaussian Noise with Standard deviation $0.8\sigma{}$ or lower reduces the residuals out of the AR(20) model causing the CUSUM algorithm to accept the sensor evolution as legitimate although the reported sensor value does not follow a feasible temporal evolution of the physical process.

\begin{table}[tb]
\setlength{\tabcolsep}{5.5pt}
    \centering
\caption{Resiliency analysis, Urbina et al. Batadal test dataset 1, unconstrained attack}
\begin{tabular}{rccccc}
     \toprule
    Dataset & Acc. & F1 & Prec. & Rec. & FPR \\
    \midrule
        Original & 0.9150 & 0.4133 & 0.7852 & 0.2805 & 0.0092 \\ 
        Replay & 0.8893 & 0.0087 & 0.0989 & \textbf{0.0046} & 0.0050 \\ 
        Constant & 0.8892 & NaN & 0.0000 & \textbf{0.0000} & 0.0046 \\
        G.N. $\sigma$  & 0.9214 & 0.4544 & 0.8790 & 0.3064 & 0.0050 \\
        G.N. $0.9\sigma$& 0.9020 & 0.2086 & 0.7580 & 0.1209 & 0.0046 \\
        G.N. $0.8\sigma$& 0.8948 & 0.0977 & 0.5801 & \textbf{0.0534} & 0.0046 \\ 
        G.N. $0.6\sigma$ & 0.8912 & 0.0356 & 0.3304 & \textbf{0.0188} & 0.0046 \\  
        G.N. $0.5\sigma$ & 0.8896 & 0.0088 & 0.1071 & \textbf{0.0046} & 0.0046 \\ 
    \bottomrule
\end{tabular} 
\label{tab:results_urbina}
    \label{tab:my_label}
\end{table}

\subsection{Chen et al.}

\begin{table}[t]
\setlength{\tabcolsep}{5.5pt}
    \centering
    \caption{Resiliency analysis, Chen et al.}
    \begin{tabular}{rccccc}
    \toprule
    Dataset & Acc. & F1 & Prec. & Rec. & FPR \\
    \midrule
    Original& 0.98 & 0.98 & 0.96 & 1.00 & 0.04 \\
    Constant Mean & 0.98 & 0.98 & 0.96 & 1.00 & 0.04 \\
    Constant Median & 0.98 & 0.98 & 0.96 & 1.00 & 0.04 \\
    Unconstrained Gaussian & 0.98 & 0.98 & 0.96 & 1.00 & 0.04 \\
    Unconstrained Replay & 0.48 & 0.01 & 0.09 & \textbf{0.00} & 0.04 \\
    Constrained Replay 1/3 & 0.96 & 0.96 & 0.96 & 0.96 & 0.04 \\
    Constrained Replay 2/3 & 0.86 & 0.85 & 0.95 & \textbf{0.77} & 0.04 \\
    \bottomrule
    \end{tabular}
    \label{tab:Chen}
\end{table}

We target the SVM detector with our \up attacks. In Table~\ref{tab:Chen} we report the results of our analysis. This is the first detector that is resilient against our \up manipulations. In particular, we found that our Constant and Gaussian attacks completely fail in their purpose. This result comes from the fact that the SVM captures both the Spatial and Temporal correlation of the sensor readings. The only attack that succeeds is the Unconstrained Replay Attack. Replay attack remains undetected because it spoofs sensor readings with benign values. Moreover the detector is not able to detect the start and end of an attack because the temporal consistency is checked with a delay of $100$ time steps constrained Replay attack reduces the recall to 0.77 when the attacker can spoof 2 out of 3 features. This result confirms that an attacker has to deal with both temporal and spatial consistency to create a feature vector that is classified benign by the SVM. 

In literature techniques to craft adversarial examples against SVM models~\cite{biggio13evasion} were proposed. Those techniques are out of scope in our study as they require to have white-box access to the anomaly detector (or to build a surrogate model) and cannot be precomputed.

\Par{Further investigations} We further investigate the root of the detector's robustness with a twofold objective. (i) Determine how Spatial and Temporal characteristics of the anomaly detector influence its robustness against \up. (ii) Determine if the different approach used in the training phase (training on both benign and abnormal sensor readings) impacts the model robustness. Due to space constraints, complete results can be found in Appendix~\ref{app:chen}.

To address (i), we remove one by one the spatial and temporal characteristics and re-train the SVM detector on a subset of features. If we remove the temporal features and train on ($\pi$), the attacker has to spoof sensor values in a spatial-consistent way, spoofing the median of the benign data without knowing the physics of the process brings Recall from $0.99$ to $0.0$, while the detector is resilient to the other manipulations (See Appendix Table~\ref{tab:Chen2}). If we remove the spatial features, i.e.,~train on a single sensor reading $i$ at time $t$ and $t+100$ ($\pi_i, \pi'_i$) the attacker has to spoof values in a temporal-consistent way. All our \up attacks succeed in the evasion bringing the Recall score to $0.0$ (See Appendix Table~\ref{tab:Chen3}). Finally, we train the SVM on univariate sensor values without the temporal evolution, i.e.,~($\pi_i$). Again all our \up succeed in the classifier evasion despite the original detector performance (See Appendix Table~\ref{tab:Chen4}). 

To address (ii), we train an OneClass SVM on ($\pi$,$\pi'$) benign data instead of training with both benign and anomalous sensors. The model is robust to our manipulations as occurred in the original case, demonstrating the independence of robustness w.r.t. training samples (See Appendix Table~\ref{tab:Chen5}).

Our results demonstrate that the robustness of the SVM model proposed by Chen et al. is given by accounting for Spatial and Temporal properties of sensor readings. Removing those features the trained models despite a good original detection performance are not capable anymore to correctly capture the physics of the process becoming vulnerable to \up attacks.

\subsection{Summary of Findings}
Table~\ref{tab:summary} reports a summary of the results of our experiments. The table is divided into two sections. The first reports the expected properties to be verified by the anomaly detector. The second reports the actual vulnerabilities that the detector was found susceptible to after testing with our framework. We can observe that a number of attacks were successful although we expected the detector to be resilient to them.

\begin{table}[tb]
\setlength{\tabcolsep}{3pt}
\caption{Summary of expected and actual resilience to \up attacks. We report the properties of detectors as Expected and the Actual Resilience. 
The properties are divided according to: Spatial consistency (Sp.), Temporal consistency (Tp.) and Statistical properties (St.).
}
\begin{center}

\begin{tabular}{r ccc ccc} \toprule
                                         &    \multicolumn{3}{c}{Expected}   & \multicolumn{3}{c}{Actual} \\ %
                                         &\multicolumn{3}{c}{Resilience}   &\multicolumn{3}{c}{Resilience}      \\ \cmidrule(l){2-4}\cmidrule(l){5-7}
Detector                                 &    Sp.    & Te. &  St.&    Sp.    & Te. &  St. \\
\cmidrule(l){1-1} \cmidrule(l){2-4} \cmidrule(l){5-7}
Aoudi et al.~\cite{aoudi18truth}         &     \xmark   &  \cmark   &   \tild &    \xmark   &  \xmark   &   \xmark     \\
Feng et al.~\cite{feng2019systematic}    &    \cmark   &  \cmark &   \tild &  \xmark   &  \tild  &   \xmark      \\
Urbina et al.~\cite{urbina16limiting}&   \xmark   &  \cmark &   \tild &  \xmark   &  \xmark  &   \xmark \\
Chen et al.~\cite{chen2018learning}&   \cmark   &  \cmark &   \tild &  \cmark   &  \tild  &   \cmark \\
\bottomrule
\end{tabular}

\label{tab:summary}
\end{center}
\end{table}

For Aoudi et al~\cite{aoudi18truth} and Urbina et al.~\cite{urbina16limiting}, after our analysis, the detectors were expected to be resilient against attacks breaking Temporal consistency. Our experiments surprisingly show how an attacker leveraging on Gaussian Noise sensor spoofing (that breaks temporal consistency) can avoid detection. For Feng et al.~\cite{feng2019systematic}, after our analysis, the detector was expected to be resilient against attacks breaking spatial and temporal consistencies. Our results surprisingly show that the anomaly detector does not trigger alarms when Spatial consistency is broken, and it can be partially considered vulnerable to attacks that break Temporal consistency. For Chen et al.~\cite{chen2018learning}, after our analysis, the detector was found resilient against attacks breaking spatial and temporal consistencies. We investigated the root of this robustness against \up, and we found that it lies in the combination of both Spatial and Temporal features analyzed by the SVM.

The analysis and the results found in our contribution highlight the need for new datasets to evaluate the performance of anomaly detectors in ICS. Our proposed attacks (which we consider to be well in the attacker model) were not appropriately represented in the datasets, and the prior work was likely also not designed to detect them. As such, we see this work as encouragement of discussion on the resiliency of anomaly detectors when analyzed against targeted manipulations.

Our proposed \up, although simple and easy to spot by a human operator, are effective in concealing the true state of the system to the anomaly detector. We believe that a critical analysis of the results, beyond good performances over metrics, is the way to pursue security of ICS. Towards this aim, our contribution proposes a framework to systematically assess detector performance under three important properties (i.e.,~Spatial consistency, temporal Consistency and Statistical properties) for effective anomaly detection. We recommend enhancing current attack datasets with the \up attacks we presented in this work, to guide the design of anomaly detectors towards a class of efficient and robust detectors.

\section{Related Work}
\label{sec:related}
Note: anomaly detection work is discussed in Section~\ref{sec:properties}. 

Adversarial Machine Learning is the research topic at the intersection of Machine Learning and System Security, this field investigates the security properties of machine learning algorithms when targeted by attackers. Attackers in the Adversarial Machine Learning setting can be motivated to perform a different type of attacks~\cite{huang2011adversarial}. In particular: classifier evasion, model poisoning, model stealing, inference attacks. Several works have been presented in the field of image classification.

Classifier evasion in the field of Cyber-Physical System is a rising research topic. So far, attacks that target Deep Learning-based classifiers have been proposed. The work~\cite{feng2017deep} proposes the usage of Generative Adversarial Networks to produce black-box stealthy manipulations and fool LSTM based defense for Industrial Control Systems. In~\cite{erba2020concealment}, two real-time evasion attacks against reconstruction-based classifiers are proposed. A black-box and a white-box attack method are presented. This work is the first that models an attacker in the setting of Adversarial Machine Learning for ICS.  In~\cite{kravchik2019efficient}, an anomaly detector based on Auto-encoders is presented, the attacker from~\cite{erba2020concealment} is weakened to demonstrate robustness to adversarial examples for the presented anomaly detector. An iterative gradient-based attack is proposed. The work~\cite{zizzo2019intrusion} evaluates the impact of adversarial examples in ICS by applying white box FGSM~\cite{goodfellow14explainingfgsm} to LSTM anomaly detectors.

\section{Conclusions}
\label{sec:conclusion}

In this work, we provided the first systematic analysis of the resilience of model-free process-based anomaly detection schemes, and introduced a taxonomy of properties that are verified (i.e.,~temporal, spatial, and statistical features). We argued that all reviewed schemes (at least implicitly) base their classification on a subset of those three properties. 
We presented a general framework to synthetically spoof those properties to hide anomalies from the detectors. Our attacks only rely on (passive) observations of normal system operations and do not require white-box access to the classifiers. 
In addition, our attacks do not require physical process knowledge (previously required for stealthy attacks). 

Using our framework, we tested \num recent model-free anomaly detectors from prior work~\cite{aoudi18truth,feng2019systematic, urbina16limiting, chen2018learning} whose security properties were not investigated yet (requiring re-implementation the classifiers in~\cite{feng2019systematic, urbina16limiting}). Three of the analyzed detectors are susceptible to most of our \up. Our attacks reduced the Recall of~\cite{aoudi18truth} from $0.63$ to $0.06$, of~\cite{feng2019systematic} from $0.47$ to $0.0$, and of ~\cite{urbina16limiting} from $0.28$ to $0.0$. The detector from Chen et al.~\cite{chen2018learning} is resilient against our \up attacks, unless the attacker performs an unconstrained Replay. Moreover, we demonstrated that detectors become vulnerable to our attacks when removing temporal and spatial features from the model.

The weaknesses we demonstrated in the anomaly detectors show that (despite good detection performance of the original schemes), the detectors are not able to reliably learn process physics and detect adversarially manipulated physical system properties unless they account for both spatial and temporal correlations. Even attacks that prior work was expected to be resilient against (based on verified properties) were found to be successful (e.g.,~Constant spoofing). We argue that our findings demonstrate the need for both more complete attacks in datasets, and critical analysis of process-based detectors.

We plan to release our framework together with the \up data generated.

\section{Acknowledgments}
 We thank the authors of~\cite{chen2018learning} for providing us with the SVM detector and data used in their work.

\bibliographystyle{plain}
\bibliography{main}

\begin{thebibliography}{10}

\bibitem{adepu2016distributed}
Sridhar Adepu and Aditya Mathur.
\newblock Distributed detection of single-stage multipoint cyber attacks in a
  water treatment plant.
\newblock In {\em Proceedings of the ACM ASIA Conference on Computer and
  Communications Security (ASIACCS)}, pages 449--460. ACM, 2016.

\bibitem{ahmed18noiseprint}
Chuadhry~Mujeeb Ahmed, Martin Ochoa, Jianying Zhou, Aditya~P. Mathur, Rizwan
  Qadeer, Carlos Murguia, and Justin Ruths.
\newblock Noiseprint: Attack detection using sensor and process noise
  fingerprint in cyber physical systems.
\newblock In {\em Proceedings of the 2018 on Asia Conference on Computer and
  Communications Security}, ASIACCS '18, pages 483--497, New York, NY, USA,
  2018. ACM.

\bibitem{aoudi18truth}
Wissam Aoudi, Mikel Iturbe, and Magnus Almgren.
\newblock Truth will out: Departure-based process-level detection of stealthy
  attacks on control systems.
\newblock In {\em Proc. of the ACM Conference on Computer and Communications
  Security (CCS)}, CCS '18, pages 817--831, New York, NY, USA, 2018. ACM.

\bibitem{biggio13evasion}
Battista Biggio, Igino Corona, Davide Maiorca, Blaine Nelson, Nedim
  {\v{S}}rndi{\'{c}}, Pavel Laskov, Giorgio Giacinto, and Fabio Roli.
\newblock Evasion attacks against machine learning at test time.
\newblock In Hendrik Blockeel, Kristian Kersting, Siegfried Nijssen, and Filip
  {\v{Z}}elezn{\'y}, editors, {\em Machine Learning and Knowledge Discovery in
  Databases}, pages 387--402, 2013.

\bibitem{bristeau2010hardware}
P-J Bristeau, Eric Dorveaux, David Vissi{\`e}re, and Nicolas Petit.
\newblock Hardware and software architecture for state estimation on an
  experimental low-cost small-scaled helicopter.
\newblock {\em Control Engineering Practice}, 18(7):733--746, 2010.

\bibitem{brockwell1991time}
Peter~J Brockwell, Richard~A Davis, and Stephen~E Fienberg.
\newblock {\em Time series: theory and methods: theory and methods}.
\newblock Springer Science \& Business Media, 1991.

\bibitem{Chen98linear}
Chi-Tsong Chen.
\newblock {\em Linear System Theory and Design}.
\newblock Oxford University Press, Inc., USA, 3rd edition, 1998.

\bibitem{chen2018learning}
Yuqi Chen, Christopher~M Poskitt, and Jun Sun.
\newblock Learning from mutants: Using code mutation to learn and monitor
  invariants of a cyber-physical system.
\newblock In {\em Proc. of the IEEE Symposium on Security and Privacy}, pages
  648--660. IEEE, 2018.

\bibitem{choi2018detecting}
Hongjun Choi, Wen-Chuan Lee, Yousra Aafer, Fan Fei, Zhan Tu, Xiangyu Zhang,
  Dongyan Xu, and Xinyan Xinyan.
\newblock Detecting attacks against robotic vehicles: A control invariant
  approach.
\newblock In {\em Proc. of the ACM Conference on Computer and Communications
  Security (CCS)}, pages 801--816. ACM, 2018.

\bibitem{dash2019out}
Pritam Dash, Mehdi Karimibiuki, and Karthik Pattabiraman.
\newblock Out of control: stealthy attacks against robotic vehicles protected
  by control-based techniques.
\newblock In {\em Proceedings of the 35th Annual Computer Security Applications
  Conference}, pages 660--672, 2019.

\bibitem{erba2020concealment}
Alessandro Erba, Riccardo Taormina, Stefano Galelli, Marcello Pogliani, Michele
  Carminati, Stefano Zanero, and Nils~Ole Tippenhauer.
\newblock Constrained concealment attacks against reconstruction-based anomaly
  detectors in industrial control systems.
\newblock In {\em Proceedings of the Annual Computer Security Applications
  Conference (ACSAC)}, 2020.

\bibitem{feng2017deep}
Cheng Feng, Tingting Li, Zhanxing Zhu, and Deeph Chana.
\newblock A deep learning-based framework for conducting stealthy attacks in
  industrial control systems.
\newblock {\em arXiv preprint arXiv:1709.06397}, 2017.

\bibitem{feng2019systematic}
Cheng Feng, Venkata~Reddy Palleti, Aditya Mathur, and Deeph Chana.
\newblock A systematic framework to generate invariants for anomaly detection
  in industrial control systems.
\newblock In {\em Proc. Network and Distributed System Security Symp. (NDSS)},
  2019.

\bibitem{galloway2013introduction}
Brendan Galloway, Gerhard~P Hancke, et~al.
\newblock Introduction to industrial control networks.
\newblock {\em IEEE Communications Surveys and Tutorials}, 15(2):860--880,
  2013.

\bibitem{garcia17hey}
Luis Garcia, Ferdinand Brasser, Mehmet~H. Cintuglu, Ahmad-Reza Sadeghi, Osama
  Mohammed, and Saman~A. Zonouz.
\newblock Hey, my malware knows physics! attacking plcs with physical model
  aware rootkit.
\newblock In {\em Proceedings of the Annual Network {\&} Distributed System
  Security Symposium (NDSS)}, February 2017.

\bibitem{goh2016dataset}
Jonathan Goh, Sridhar Adepu, Khurum~Nazir Junejo, and Aditya Mathur.
\newblock A dataset to support research in the design of secure water treatment
  systems.
\newblock In {\em International Conference on Critical Information
  Infrastructures Security (CRITIS)}, pages 88--99. Springer, 2016.

\bibitem{goh2017anomaly}
Jonathan Goh, Sridhar Adepu, Marcus Tan, and Zi~Shan Lee.
\newblock Anomaly detection in cyber physical systems using recurrent neural
  networks.
\newblock In {\em High Assurance Systems Engineering (HASE), 2017 IEEE 18th
  International Symposium on}, pages 140--145. IEEE, 2017.

\bibitem{goodfellow14explainingfgsm}
Ian~J. Goodfellow, Jonathon Shlens, and Christian Szegedy.
\newblock Explaining and harnessing adversarial examples.
\newblock {\em CoRR}, abs/1412.6572, 2014.

\bibitem{hau2019exploiting}
Zhongyuan Hau and Emil~C Lupu.
\newblock Exploiting correlations to detect false data injections in
  low-density wireless sensor networks.
\newblock In {\em Proceedings of the 5th on Cyber-Physical System Security
  Workshop}, pages 1--12. ACM, 2019.

\bibitem{huang2011adversarial}
Ling Huang, Anthony~D Joseph, Blaine Nelson, Benjamin~IP Rubinstein, and
  JD~Tygar.
\newblock Adversarial machine learning.
\newblock In {\em Proceedings of the 4th ACM workshop on Security and
  artificial intelligence}, pages 43--58. ACM, 2011.

\bibitem{illiano2015detecting}
Vittorio~P Illiano and Emil~C Lupu.
\newblock Detecting malicious data injections in wireless sensor networks: A
  survey.
\newblock {\em ACM Computing Surveys (CSUR)}, 48(2):24, 2015.

\bibitem{wadi17dataset}
{iTrust, Centre for Research in Cyber Security, Singapore University of
  Technology and Design}.
\newblock {WADI datatset}, 2017.
\newblock
  {\url{https://itrust.sutd.edu.sg/research/dataset/dataset_characteristics/\#wadi}},
  Last accessed on: 2019-01-30.

\bibitem{kiran2011cfp-growth++}
R~Uday Kiran and P~Krishna Reddy.
\newblock Novel techniques to reduce search space in multiple minimum
  supports-based frequent pattern mining algorithms.
\newblock In {\em Proceedings of the 14th international conference on extending
  database technology}, pages 11--20. ACM, 2011.

\bibitem{kravchik2018detecting}
Moshe Kravchik and Asaf Shabtai.
\newblock Detecting cyber attacks in industrial control systems using
  convolutional neural networks.
\newblock In {\em Proceedings of the 2018 Workshop on Cyber-Physical Systems
  Security and PrivaCy}, pages 72--83. ACM, 2018.

\bibitem{kravchik2019efficient}
Moshe Kravchik and Asaf Shabtai.
\newblock Efficient cyber attacks detection in industrial control systems using
  lightweight neural networks.
\newblock {\em arXiv preprint arXiv:1907.01216}, 2019.

\bibitem{kune2013ghost}
Denis~Foo Kune, John Backes, Shane~S Clark, Daniel Kramer, Matthew Reynolds,
  Kevin Fu, Yongdae Kim, and Wenyuan Xu.
\newblock Ghost talk: Mitigating emi signal injection attacks against analog
  sensors.
\newblock In {\em Security and Privacy (SP), 2013 IEEE Symposium on}, pages
  145--159. IEEE, 2013.

\bibitem{mathur16swat}
Aditya Mathur and Nils~Ole Tippenhauer.
\newblock {SWaT}: A water treatment testbed for research and training on {ICS}
  security.
\newblock In {\em Proceedings of Workshop on Cyber-Physical Systems for Smart
  Water Networks (CySWater)}, April 2016.

\bibitem{moSinopoli2009}
Yilin Mo and B.~Sinopoli.
\newblock Secure control against replay attacks.
\newblock In {\em Communication, Control, and Computing, 2009. Allerton 2009.
  47th Annual Allerton Conference on}, pages 911--918, 2009.

\bibitem{mo2009secure}
Yilin Mo and Bruno Sinopoli.
\newblock Secure control against replay attacks.
\newblock In {\em Communication, Control, and Computing, 2009. Allerton 2009.
  47th Annual Allerton Conference on}, pages 911--918. IEEE, 2009.

\bibitem{ostfeld12ctown}
Avi Ostfeld, Elad Salomons, Lindell Ormsbee, James Uber, Christopher Bros, Paul
  Kalungi, Richard Burd, Boguslawa Zazula-Coetzee, Teddy Belrain, Doosun Kang,
  Kevin Lansey, Hailiang Shen, Edward McBean, Zheng Wu, Tom Walski, Stefano
  Alvisi, Marco Franchini, Joshua P.~Johnson, Santosh Ghimire, and Robert
  McKillop.
\newblock Battle of the water calibration networks.
\newblock {\em JOURNAL OF WATER RESOURCES PLANNING AND MANAGEMENT-ASCE},
  138:523--532, 09 2012.

\bibitem{quinonezsavior}
Raul Quinonez, Jairo Giraldo, Luis Salazar, Erick Bauman, Alvaro Cardenas, and
  Zhiqiang Lin.
\newblock Savior: Securing autonomous vehicles with robust physical invariants.
\newblock In {\em Proc. of the USENIX Security Symposium}, Boston, MA, August
  2020.

\bibitem{shen2020drift}
Junjie Shen, Jun~Yeon Won, Zeyuan Chen, and Qi~Alfred Chen.
\newblock Drift with devil: Security of multi-sensor fusion based localization
  in high-level autonomous driving under {GPS} spoofing.
\newblock In {\em 29th {USENIX} Security Symposium ({USENIX} Security 20)},
  pages 931--948. {USENIX} Association, August 2020.

\bibitem{taormina2019epanetCPA}
R.~Taormina, S.~Galelli, H.C. Douglas, N.~O. Tippenhauer, E.~Salomons, and
  A.~Ostfeld.
\newblock A toolbox for assessing the impacts of cyber-physical attacks on
  water distribution systems. environmental modelling software.
\newblock {\em Environmental Modelling Software}, 112:46--51, 02 2019.

\bibitem{taormina2018deep}
Riccardo Taormina and Stefano Galelli.
\newblock A deep learning approach for the detection and localization of
  cyber-physical attacks on water distribution systems.
\newblock {\em Journal of Water Resources Planning Management},
  144(10):04018065, 2018.

\bibitem{taormina18battle}
Riccardo Taormina, Stefano Galelli, Nils~Ole Tippenhauer, Elad Salomons, Avi
  Ostfeld, Demetrios~G. Eliades, Mohsen Aghashahi, Raanju Sundararajan, Mohsen
  Pourahmadi, M.~Katherine Banks, B.~M. Brentan, Enrique Campbell, G.~Lima,
  D.~Manzi, D.~Ayala-Cabrera, M.~Herrera, I.~Montalvo, J.~Izquierdo,
  E.~Luvizotto, Jr, Sarin~E. Chandy, Amin Rasekh, Zachary~A. Barker, Bruce
  Campbell, M.~Ehsan Shafiee, Marcio Giacomoni, Nikolaos Gatsis, Ahmad Taha,
  Ahmed~A. Abokifa, Kelsey Haddad, Cynthia~S. Lo, Pratim Biswas, Bijay Pasha,
  M. Fayzul K.and~Kc, Saravanakumar~Lakshmanan Somasundaram, Mashor Housh, and
  Ziv Ohar.
\newblock The battle of the attack detection algorithms: Disclosing cyber
  attacks on water distribution networks.
\newblock {\em Journal of Water Resources Planning and Management}, 144(8),
  August 2018.

\bibitem{tu19trick}
Yazhou Tu, Sara Rampazzi, Bin Hao, Angel Rodriguez, Kevin Fu, and Xiali Hei.
\newblock Trick or heat? manipulating critical temperature-based control
  systems using rectification attacks.
\newblock In {\em Proceedings of the 2019 ACM SIGSAC Conference on Computer and
  Communications Security}, CCS ’19, page 2301–2315, New York, NY, USA,
  2019. Association for Computing Machinery.

\bibitem{urbina16limiting}
David Urbina, Jairo Giraldo, Alvaro~A. Cardenas, Nils~Ole Tippenhauer, Junia
  Valente, Mustafa Faisal, Justin Ruths, Richard Candell, and Henrik Sandberg.
\newblock Limiting the impact of stealthy attacks on industrial control
  systems.
\newblock In {\em Proc. of the ACM Conference on Computer and Communications
  Security (CCS)}, October 2016.

\bibitem{urbina16fieldbus}
David Urbina, Jairo Giraldo, Nils~Ole Tippenhauer, and Alvaro C\'ardenas.
\newblock Attacking fieldbus communications in {ICS}: Applications to the
  {SWaT} testbed.
\newblock In {\em Proceedings of Singapore Cyber Security Conference (SG-CRC)},
  January 2016.

\bibitem{weinbergerStuxnet}
Sharon Weinberger.
\newblock Computer security: Is this the start of cyberwarfare?
\newblock {\em Nature}, 174:142--145, June 2011.

\bibitem{yan2016can}
Chen Yan, Wenyuan Xu, and Jianhao Liu.
\newblock Can you trust autonomous vehicles: Contactless attacks against
  sensors of self-driving vehicle.
\newblock {\em DEF CON}, 24, 2016.

\bibitem{zizzo2019intrusion}
Giulio Zizzo, Chris Hankin, Sergio Maffeis, and Kevin Jones.
\newblock Intrusion detection for industrial control systems: Evaluation
  analysis and adversarial attacks.
\newblock {\em arXiv preprint arXiv:1911.04278}, 2019.

\end{thebibliography}

\appendices
\section{Methodology}
\label{sec:methodology}
In order to evaluate the performance of the anomaly detector we observe how Accuracy Eq.~\ref{eq:accuracy}, Precision Eq.~\ref{eq:precision}, Recall Eq.~\ref{eq:recall}, and False Positive Rate Eq.~\ref{eq:falsepositiverate} scores change when the spoofing technique is applied to the data. 
\begin{equation}
\text{Accuracy} = \frac{\text{TP}+\text{TN}}{\text{TP}+\text{FP}+\text{TN}+\text{FN}}
\label{eq:accuracy}
\end{equation}

\begin{equation}
\text{Precision} = \frac{\text{TP}}{\text{TP}+\text{FP}}
\label{eq:precision}
\end{equation}

\begin{equation}
\text{Recall} = \frac{\text{TP}}{\text{TP}+\text{FN}}
\label{eq:recall}
\end{equation}

\begin{equation}
\text{F1-Score} = 2\times\frac{\text{Precision}\times \text{Recall}}{\text{Precision}+\text{Recall}}
\label{eq:f1score}
\end{equation}

\begin{equation}
\text{FPR} = \frac{\text{FP}}{\text{TN}+\text{FP}}
\label{eq:falsepositiverate}
\end{equation}

Given the original classification scores (e.g.,~when no spoofing is applied to data), \up is effective if Precision and Recall score reduces substantially. When those two scores reduce, towards $0$, it means that the instances where the \up was applied were misclassified moving them from being True Positives to False Negatives. Looking at False Positive Rate (FPR) score we can also verify if the attacks are introducing False Positive in the Classification. If the FPR remains almost like the original, it means that the \up did not induce any wrong classification (as expected since we are not spoofing data outside the boundaries of the attacks present in the dataset). Finally, since the datasets are unbalanced, with more samples with of the negative class, Accuracy score will not reach zero but at most the baseline where all the instances are labeled as the negative class.

\section{Aoudi et al. Qualitative Evaluation}
\label{sec:qualitative}
\begin{figure*}
    \centering
    \subfigure[]{\includegraphics[scale = 0.17]{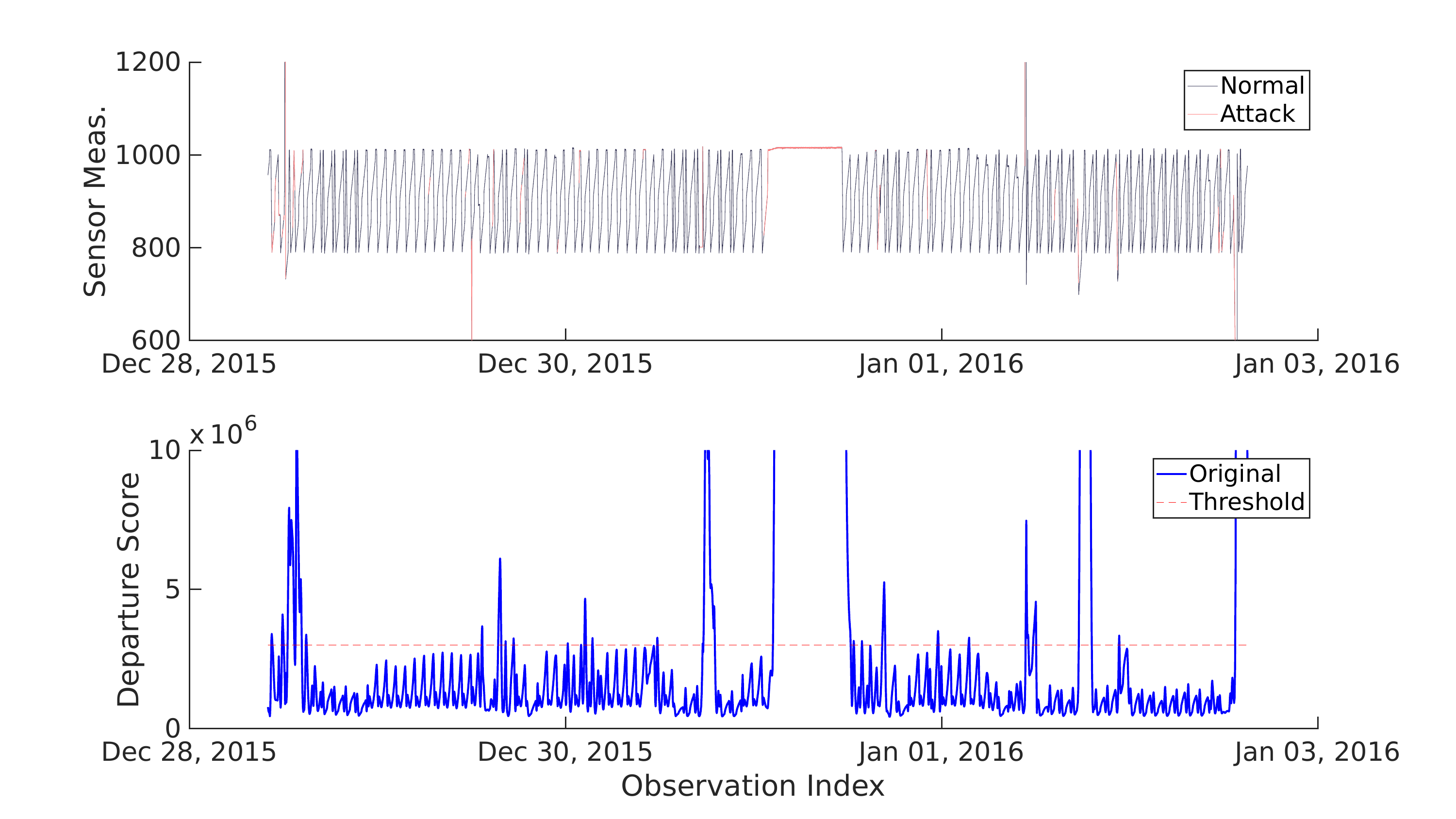}}
    \subfigure[]{\includegraphics[scale = 0.17]{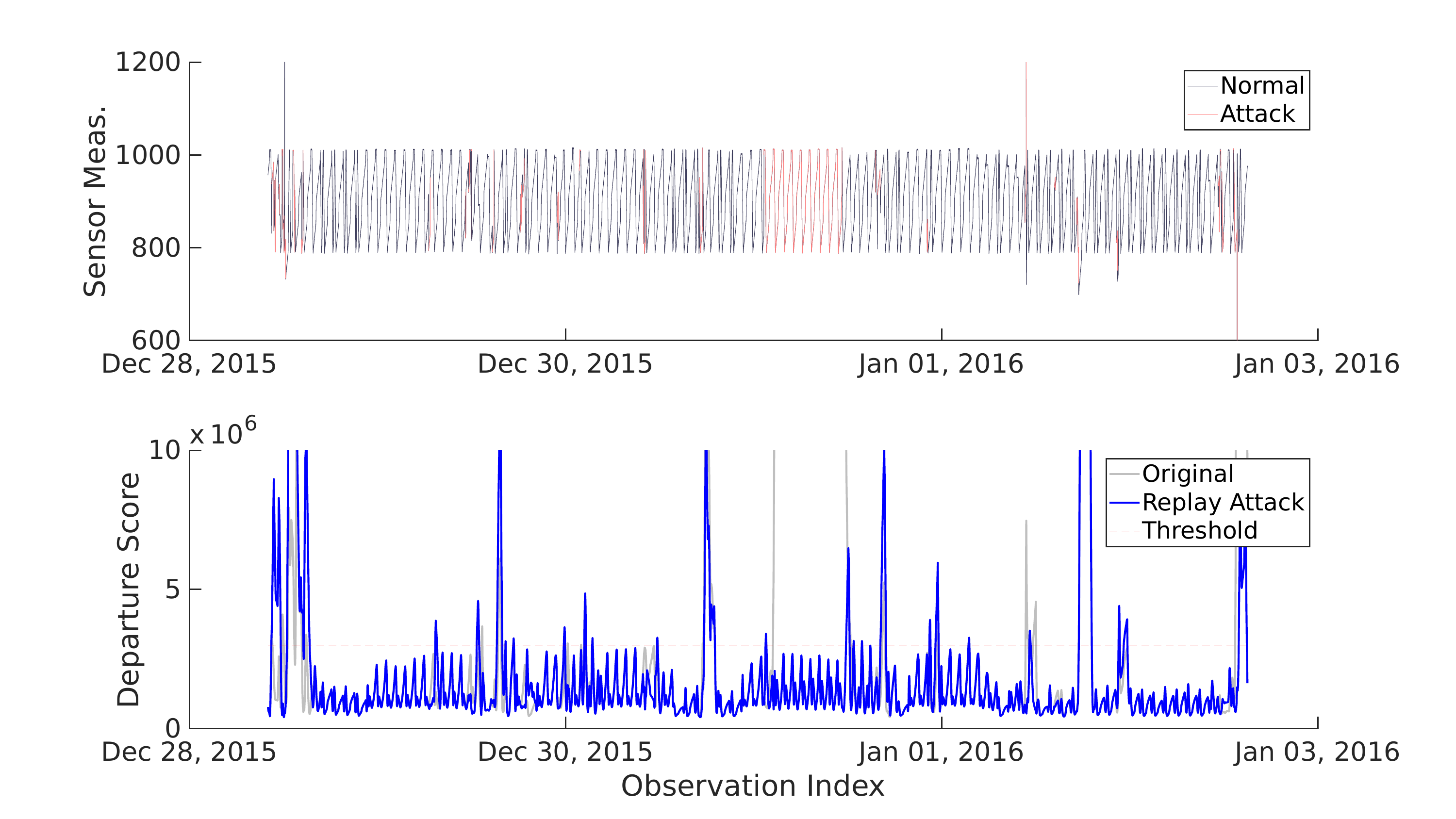}}
    \subfigure[]{\includegraphics[scale = 0.17]{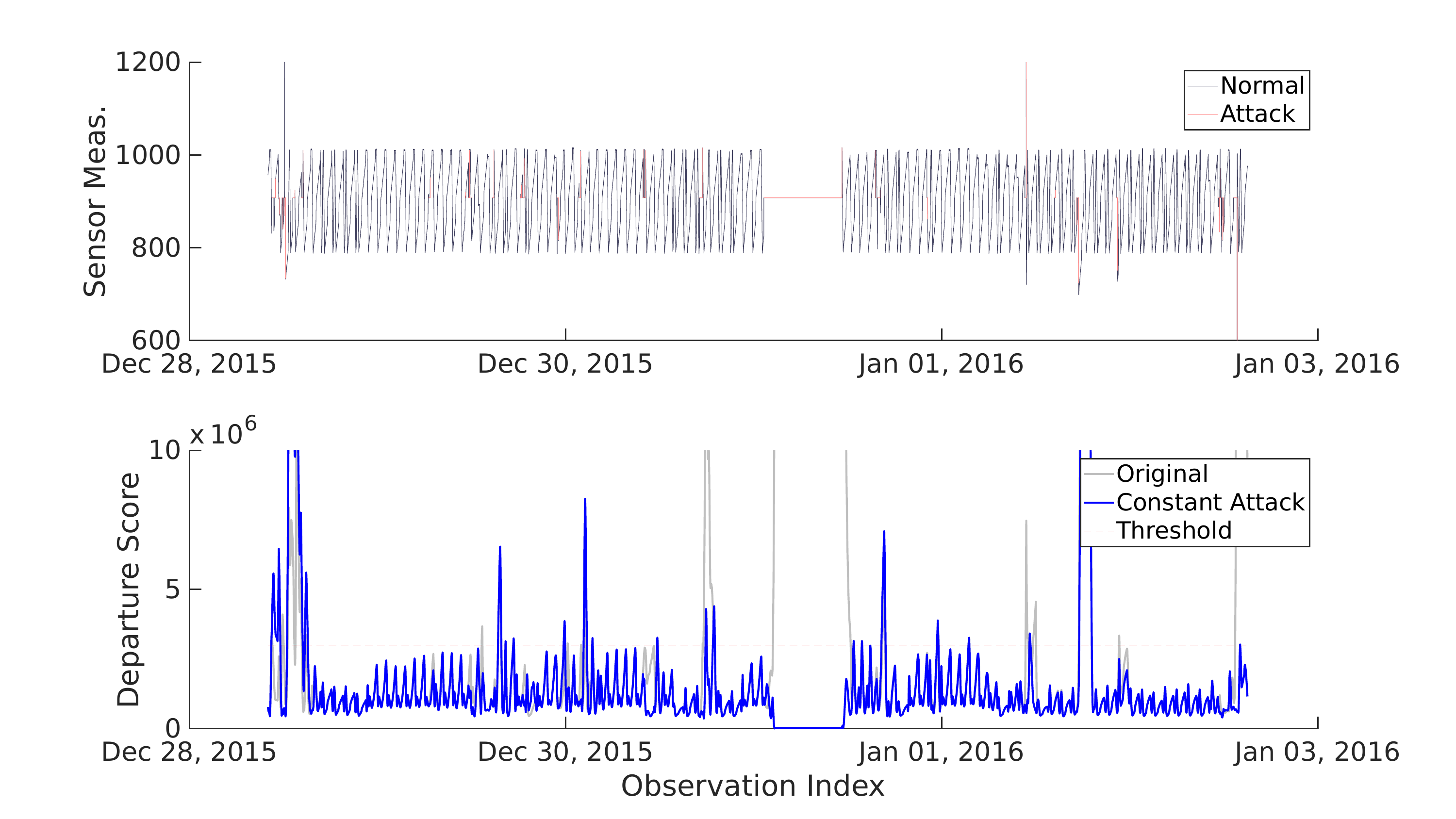}}
    \subfigure[]{\includegraphics[scale = 0.17]{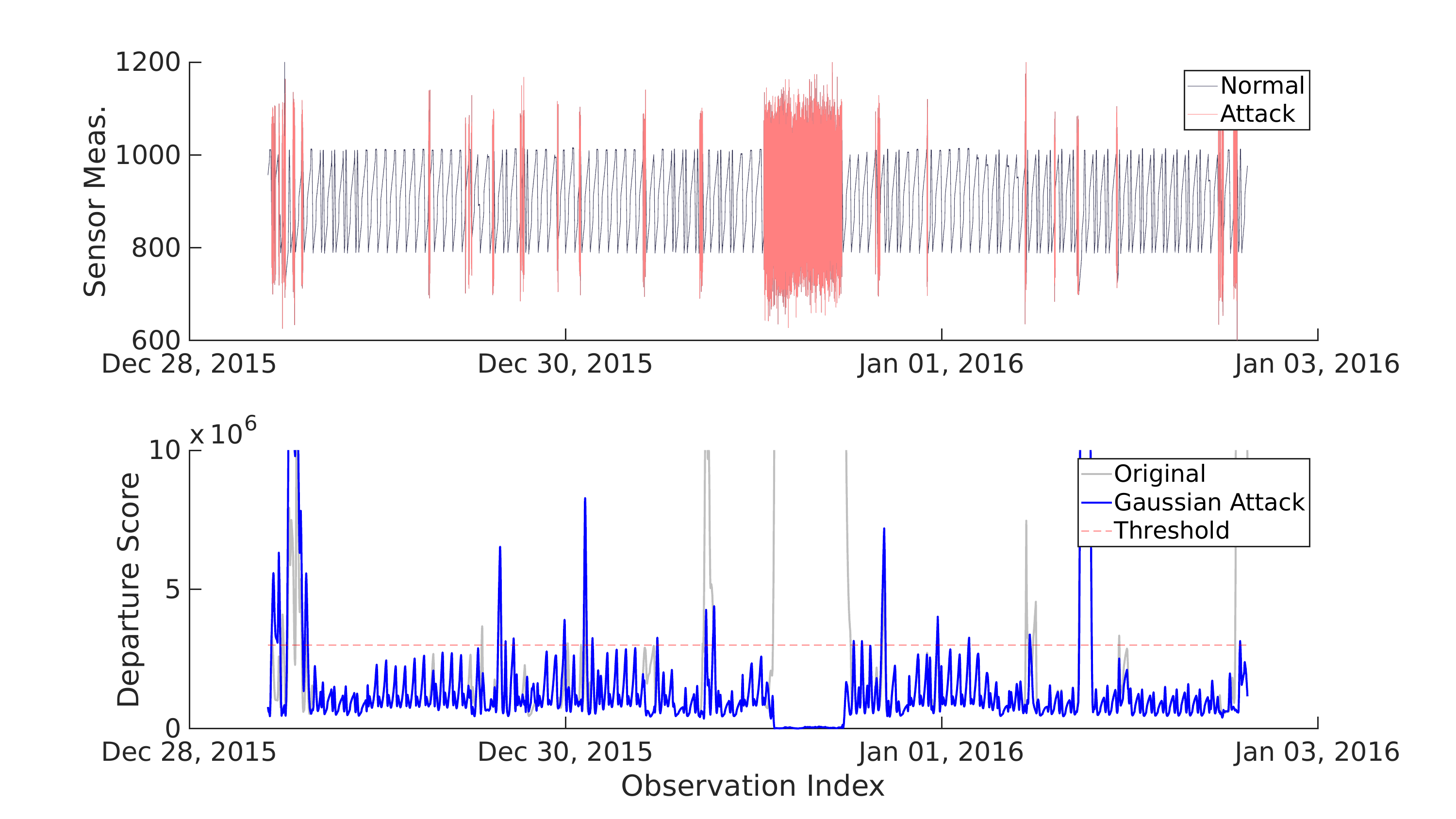}}
    
    \caption{Plots of \up attacks against Aoudi et al. defense. Test is done over sensor LIT301 of SWaT dataset (a) Original, (b) Replay attack, (c) Constant, (d) Gaussian, Threshold $3\times10^6$. Red sensor measurements are indicating values during an attack. Attacks are detected when the departure score is above the red threshold (taken from~\cite{aoudi18truth}). Our Constant and Gaussian attack lead to zero departure score for the attack on December 31 at 1 am.}
    \label{fig:aoudi_result}
\end{figure*}

Like in the evaluation from the original paper, we show sensor LIT301 from SWaT dataset. In Figure~\ref{fig:aoudi_result}(a), the departure score obtained over original SWaT reflects the one presented in~\cite{aoudi18truth}. We can notice that for the long attack (starting after 2), the departure score surpasses $10x10^6$, for evaluation purposes this attack is taken as reference. In Figure~\ref{fig:aoudi_result}(b) we can see how the departure score is influenced when the Replay Attack is applied. The reference attack has now Departure score lower than the threshold $3 x 10^6$. Figure~\ref{fig:aoudi_result}(c) shows the effect of Constant Attack. Again, we can notice that the reference attack has now Departure score close to $0$. Finally, the Gaussian Noise attack result is depicted in Figure~\ref{fig:aoudi_result}(d). As before, the Departure score related to the reference attack is close to $0$. 

\begin{table*}[tbh]
\caption{Resiliency analysis, Feng et al. Batadal test dataset 1, constrained attack. This table shows how the recall score decreases when constrained \up are performed. }
\begin{center}
    
\begin{tabular}{lccccccccccc}
 \toprule
\# Spoofed Values & 2 & 3 & 4 & 5 & 6 & 7 & 8 & 9 & 10 & 15 & 20 \\
\midrule
Replay             
& 0.44 &                          0.36 &                          0.01 &                          0.01 &                          0.01 &                          0.02 &                          0.01 &                          0.01 &                           0.02 &                           0.01 &                           0.03 \\
Constant            
 & 0.44 &                            0.37 &                            0.01 &                            0.01 &                            0.01 &                            0.01 &                            0.01 &                            0.01 &                             0.01 &                             0.01 &                             0.02 \\
Gaussian Noise     
 & 0.53 &                            0.53 &                            0.47 &                            0.46 &                            0.46 &                            0.47 &                            0.46 &                            0.57 &                             0.59 &                             0.56 &                             0.68 \\
Gaussian Noise 2    
& 0.50 &                              0.66 &                              0.66 &                              0.66 &                              0.66 &                              0.66 &                              0.66 &                              0.75 &                               0.75 &                               0.73 &                               0.69 \\

\bottomrule
\end{tabular}

\label{tab:results_resiliency_batadal_constrained}

\end{center}
\end{table*}

\begin{figure*}
    \centering
    \subfigure[]{\includegraphics[scale = 0.6]{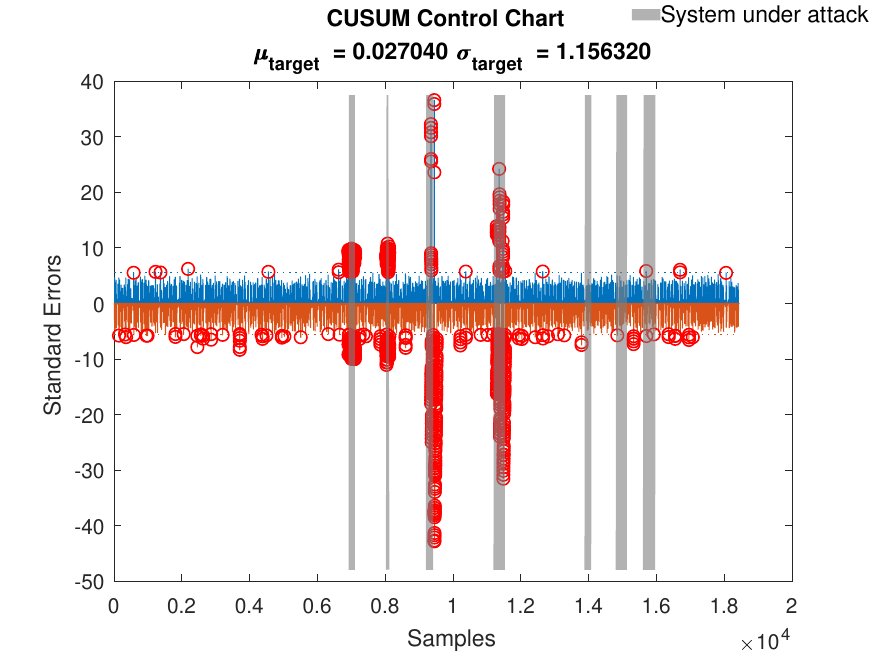}}
    \subfigure[]{\includegraphics[scale = 0.6]{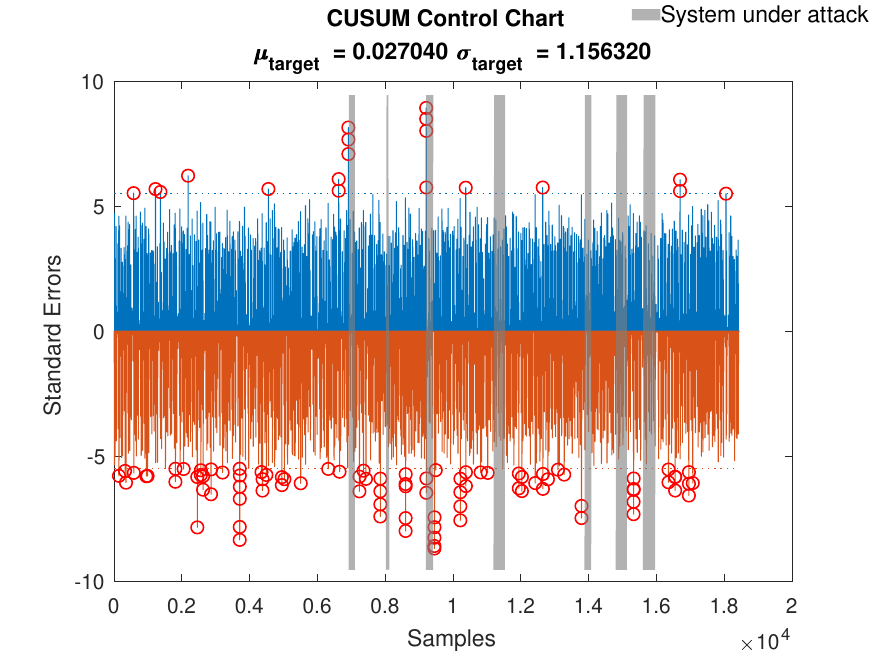}}
    \subfigure[]{\includegraphics[scale = 0.6]{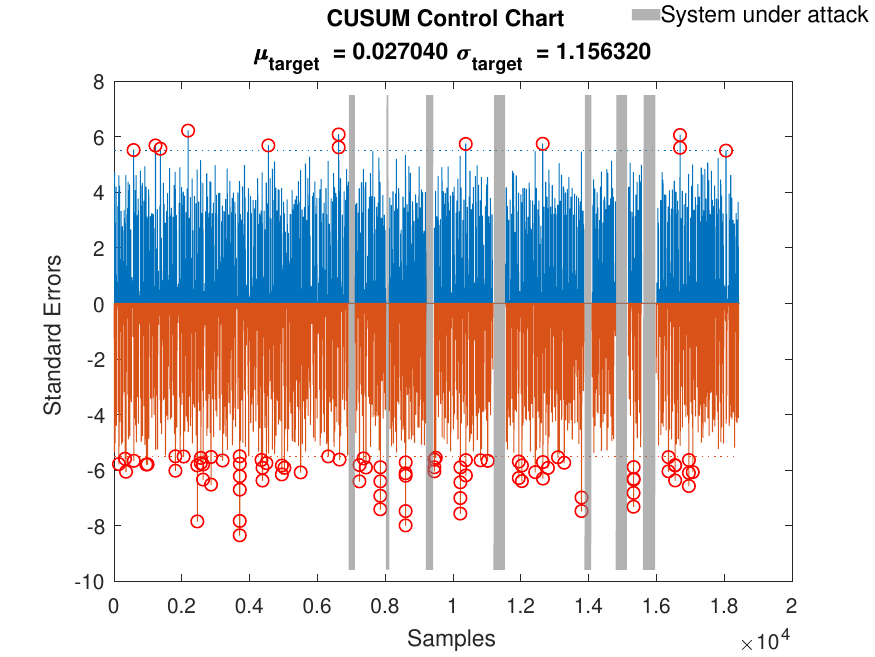}}
    \subfigure[]{\includegraphics[scale = 0.6]{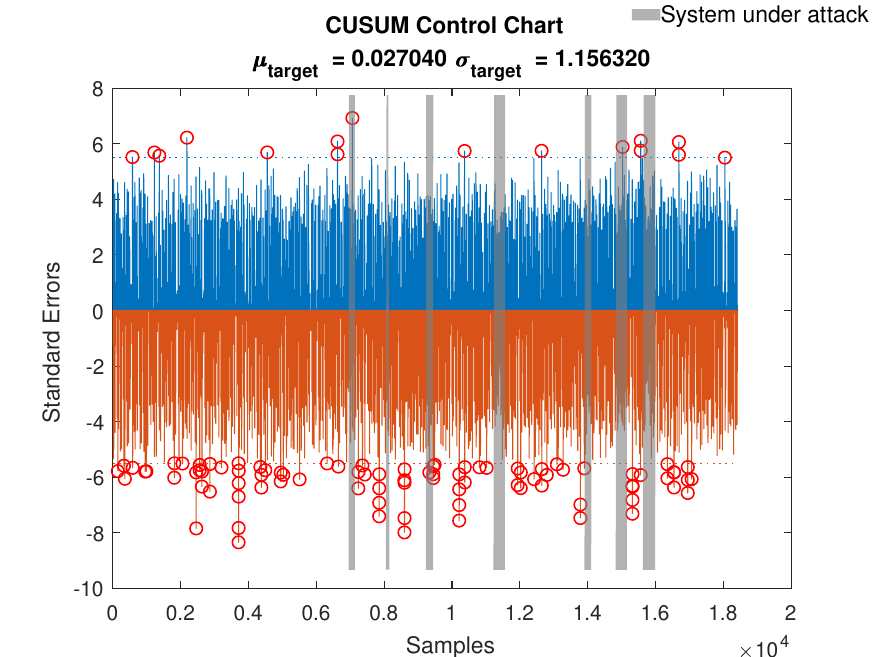}}
    
    \caption{CUSUM control chart of the \up attacks against Urbina et al. defense. Test is done over sensor PRESSURE\_J302 of BATADAL dataset (a) Original, (b) Replay attack, (c) Constant, (d) Gaussian. Red dots are indicating when the CUSUM surpasses the expected control thresholds and grey bars indicate whether an anomaly is occurring in the system. In (a) $4$ over $7$ attacks are originally detected with high confidence. All the \up in (b), (c) and (d) diminish the detector confidence hiding the anomalies from the detector, i.e.,~turning the True Positives into False Negatives. We note that the remaining red dots after applying the `up attacks are in most of the cases False Positives. }
    \label{fig:urbina_result}
\end{figure*}
\section{Chen et al. Additional Results}
\label{app:chen}
\begin{table}[t]
\setlength{\tabcolsep}{5.5pt}
    \centering
    \caption{Resiliency analysis, Chen et al. No temporal}
    \begin{tabular}{rccccc}
    \toprule
    Dataset & Acc. & F1 & Prec. & Rec. & FPR \\
    \midrule
    Original & 0.99 & 0.99 & 0.98 & 1.00 & 0.02 \\
    Constant Mean & 0.99 & 0.99 & 0.98 & 1.00 & 0.02 \\
    Constant Median & 0.49 & 0.00 & 0.00 & \textbf{0.00} & 0.02 \\
    Unconstrained Gaussian & 0.99 & 0.99 & 0.98 & 1.00 & 0.02 \\
    Unconstrained Replay & 0.49 & 0.00 & 0.09 & \textbf{0.00} & 0.02 \\
    Constrained Replay 1/3 & 0.97 & 0.97 & 0.98 & 0.97 & 0.02 \\
    Constrained Replay 2/3 & 0.88 & 0.86 & 0.98 & \textbf{0.77} & 0.02 \\
    \bottomrule
    \end{tabular}
    \label{tab:Chen2}
\end{table}

\begin{table}[t]
\setlength{\tabcolsep}{5.5pt}
    \centering
    \caption{Resiliency analysis, Chen et al. No spatial}
    \begin{tabular}{rccccc}
    \toprule
    Dataset & Acc. & F1 & Prec. & Rec. & FPR \\
    \midrule
    Original & 0.99 & 0.99 & 0.97 & 1.00 & 0.03 \\
    Constant Mean & 0.49 & 0.00 & 0.00 & \textbf{0.00} & 0.03 \\
    Constant Median & 0.49 & 0.00 & 0.00 & \textbf{0.00} & 0.03 \\
    Unconstrained Gaussian & 0.49 & 0.00 & 0.00 & \textbf{0.00} & 0.03 \\
    Unconstrained Replay & 0.50 & 0.05 & 0.49 & \textbf{0.02} & 0.03 \\
    \bottomrule
    \end{tabular}
    \label{tab:Chen3}
\end{table}

\begin{table}[t]
\setlength{\tabcolsep}{5.5pt}
    \centering
    \caption{Resiliency analysis, Chen et al. No temporal No spatial}
    \begin{tabular}{rccccc}
    \toprule
    Dataset & Acc. & F1 & Prec. & Rec. & FPR \\
    \midrule
    Original & 0.99 & 0.99 & 0.99 & 0.99 & 0.01 \\
    Constant Mean & 0.49 & 0.00 & 0.00 & \textbf{0.00} & 0.01 \\
    Constant Median & 0.49 & 0.00 & 0.00 & \textbf{0.00} & 0.01 \\
    Unconstrained Gaussian & 0.50 & 0.04 & 0.75 & \textbf{0.02} & 0.01 \\
    Unconstrained Replay & 0.50 & 0.02 & 0.56 & \textbf{0.01} & 0.01 \\
    \bottomrule
    \end{tabular}
    \label{tab:Chen4}
\end{table}

\begin{table}[t]
\setlength{\tabcolsep}{5.5pt}
    \centering
    \caption{Resiliency analysis, Chen et al. One class svm}
    \begin{tabular}{rccccc}
    \toprule
    Dataset & Acc. & F1 & Prec. & Rec. & FPR \\
    \midrule
    Original& 0.93 & 0.93 & 0.88 & 1.00 & 0.14 \\
    Constant Mean & 0.93 & 0.93 & 0.88 & 1.00 & 0.14 \\
    Constant Median & 0.93 & 0.93 & 0.88 & 1.00 & 0.14 \\
    Unconstrained Gaussian & 0.93 & 0.93 & 0.88 & 1.00 & 0.14 \\
    Unconstrained Replay & 0.48 & 0.16 & 0.41 & \textbf{0.10} & 0.14 \\
    Constrained Replay 1/3 & 0.92 & 0.92 & 0.87 & 0.97 & 0.14 \\
    Constrained Replay 2/3 & 0.83 & 0.82 & 0.85 & \textbf{0.79} & 0.14 \\
    \bottomrule
    \end{tabular}
    \label{tab:Chen5}
\end{table}

We investigate the root of the resilience of the anomaly detector among Spatial and Temporal features. To this end we removed one by one the features analyzed by the detector and test our attacks. We conducted four experiments. 

1) Detector trained without temporal features, i.e.,~only spatial feature vector ($\pi$). In Table~\ref{tab:Chen2} we report the results of the detector trained on Spatial features only. As we can see the detector results originally in an high detection performance. When spoofing the values with the median of each sensor the detector fails to spot the anomalies and the recall score drops to 0. This was not the case when accounting for both spatial an temporal consistency. 

2) We trained the detector without spatial features. i.e.,~only temporal ($\pi_i$,$\pi'_i$). In Table~\ref{tab:Chen3} we report the results of the detector trained on temporal features only. Also in this case the detector has an high detection scores. When the detector is targeted with our \up attacks, we can observe that the detector's recall drops to $0$ in all the different attacks. Removing spatial consistency makes the SVM unable to detect spoofed sensor signals because the features do not describe enough the physical properties of the system. 

3) We removed both temporal and spatial features and trained the detectors on the one dimensional feature ($\pi_i$), results are reported in Table~\ref{tab:Chen3}. As we can notice also in this case the original detector has an high detection accuracy. When testing with our attacks the detector is not capable of detecting \up that contains inconsistencies.

4) We trained a one class SVM model trained over benign samples ($\pi,\pi'$) and tested with our attack to verify that robustness of the detector depends on temporal and spatial features and not from the Two class training approach. As we can observe in Table~\ref{tab:Chen5} the detector is resilient to \up attacks. This result shows that the key of the resilience resides in correctly abstracting spatial and temporal features of the process to detect inconsistencies caused by sensor spoofing.

\end{document}